\documentclass{ptephy_v2}

\preprintnumber{KUNS-3134} 

\usepackage{ascmac}

\usepackage{xcolor}
\usepackage[normalem]{ulem}
\providecommand{\delete}[1]{} 
\providecommand{\add}[1]{#1}
\renewcommand{\comment}[1]{}

\usepackage{hyperref} 

\makeatletter
\let\ptep@subsection\l@subsection
\renewcommand*\l@subsection[2]{%
  \ifnum\c@tocdepth>1\relax
    \ptep@subsection{#1}{#2}%
  \fi
}
\let\ptep@subsubsection\l@subsubsection
\renewcommand*\l@subsubsection[2]{%
  \ifnum\c@tocdepth>2\relax
    \ptep@subsubsection{#1}{#2}%
  \fi
}
\makeatother

\begin{document}

\begin{flushright}
\vspace*{2mm}
{\small KUNS-3134} 
\end{flushright}

\title{Atom as an Agent: \\
Emergent Order by Collective Predictive Coding}


\author[1]{Koji Hashimoto}
\author[2]{Tadahiro Taniguchi}
\affil[1]{Physics Department, Kyoto University, Kyoto 606-8502, JAPAN
\email{koji@scphys.kyoto-u.ac.jp}}
\affil[2]{Graduate School of Informatics, Kyoto University, Kyoto 606-8501, JAPAN
\email{taniguchi@i.kyoto-u.ac.jp}}

%
%
%

\begin{abstract}%
What is the intrinsic reason of emergence of macroscopic order parameter out of microscopic systems? 
In this article, we apply to physics the collective predictive coding (CPC), the theory of language/symbol emergence.
As our language emerges out of our human communications, physical macroscopic order parameter emerges out of communications among atoms --- each atom is regarded as an agent who predicts each other's behavior through inference. 
We present three examples of classical spin systems: two-dimensional ferromagnetic and antiferromagnetic Ising models, a two-dimensional spin model with competing stripe orders, and the fully connected Sherrington--Kirkpatrick spin-glass model. 
In all of these examples, proper macroscopic order parameters
including the replica overlap emerge automatically through
communication informatics of CPC, only from observed spin
configurations.
We call the learned macroscopic coordinate the collective
predictive order. 
Furthermore, for the ferromagnetic and antiferromagnetic Ising
models, the equilibrium distribution of the collective predictive
order yields an emergent potential with the symmetry and thermal evolution
characteristic of a Landau potential.
Notably, the emergence of the collective predictive order does not refer to any structure of microscopic Hamiltonians/Lagrangians.
Thus our ``Atom-as-an-Agent" (AaaA) principle offers a completely data-driven method for the emergence of macroscopic order, that is, machine learning physics.

\end{abstract}


\maketitle

\setcounter{tocdepth}{1}
\tableofcontents


\section{The problem of order parameter emergence}
\label{sec:introduction}

Known is the critical importance of emergence in physics, while the reason of its existence has not been revealed.\footnote{
For a philosophical discussion of emergence and reduction in
physics, particularly the relation between microscopic and
macroscopic descriptions, see Ref.~\cite{Morita:2024Emergence}.} Numerous examples of the physical emergence include, for example, classical physics emergent from quantum physics, and classical spacetime emergent from quanta in holographic principle \cite{Maldacena:1997re,Gubser:1998bc,Witten:1998qj}. It is rather natural that the emergence is not physically explained, because the emergence is the un-explanability of collective phenomena; suppose some microscopic ingredients exhibits collectively some distinguishable property which cannot be explained by the ingredients' own  property directly, that is when it is called emergent. If this is truly the case, we may have to give up a direct derivation of, say, order parameter of the emergent phenomena, under the viewpoint that any physical derivation should be accompanied by some explainability. 

However, from a meta-level of viewpoint, the reasoning above has a pitfall. When we say we explain a certain phenomenon, we use some notions which are shared by us physicists, and the explainability actually depends on the popularity of the notions used in the community. Physical notions have been born and renewed through communication among physicists and through discovery of new phenomena in experiments and theories. Physical notions, such as the order parameters, were born as a language with which physicists communicate. Thus, this historical aspect of how physical notions are born in physics can be a hint for discovery of a possible mechanism of emergent phenomena in physics, and a possible reason of the existence of the macroscopic variables.

To this end, we are motivated to utilize an information-theoretic theory of language/symbol emergence, called Collective Predictive Coding (CPC) \cite{Taniguchi:2024CPC,Taniguchi:2023MHNG}, to find the mechanism of emergence in physics. This theory is defined with an information-theoretic inference free energy which models how languages are emergent through communication between agents.
\add{CPC was proposed as a computational account of symbol emergence systems, in which individuals develop their own perceptual representations through interactions with the environment, while shared symbol systems emerge through communication \cite{Taniguchi:2026SES}. Human society is a typical example of such a system.}\comment{CPC-01: Introduce symbol emergence systems (SES) and cite the Springer volume.} More concretely, CPC theorizes the structure of communication,
and learning to jointly explain the agents' observations leads
to the emergence of a common shared notion, which serves as
\delete{a language among the agents and enables them to predict
one another's observations.}\add{a language among the agents. Each agent predicts its own observations using its local probabilistic prediction model, while the shared variable connects these models into a collective predictive description.}\comment{CPC-02: Distinguish individual prediction of local observations from collective prediction through the shared variable.} 

As our application of CPC to physics, {\it we regard each atom as a CPC agent}. 
Here a CPC agent is a unit that uses a local probabilistic prediction model to infer a shared variable from its observations and learns the parameters of that model.
Each atom observes its state, for example its spin. \delete{From the single observation of its own spin, the atom tries to predict other atom's observation, through communication. The language needs to be emergent for this prediction to be successful.}\add{Each atom uses its local probabilistic prediction model to predict
its own spin observation from the shared variable, and participates
in inferring that variable through communication.
Through communication and learning, the shared variable acquires
a collective meaning: the local probabilistic prediction models
together predict the whole spin configuration.}\comment{CPC-03: Clarify the prediction target of each atomic agent; preserve the subsequent discussion of order parameters.} The emergent language should be nothing but the order parameter of the whole system, because the order parameter in physics is a macroscopic variable which characterizes the behavior of the whole system in a concise manner. The order parameter singles out the most essential ingredient of the system and dictates the phase structure of the system, and sharing this order parameter value among agents is the most efficient way to reproduce the whole system behavior. Thus, CPC is the key to the emergence of the order parameter.

We call this the Atom-as-an-Agent (AaaA) principle.
In this paper, we implement this principle using CPC. If this actually works, from just the observation of each atom's spin configuration, \delete{maximizing the predictability of other atoms' state}\add{improving the collective predictability of the observed spin configurations}\comment{CPC-04: Keep the AaaA description consistent with individual and collective prediction.} should lead to the automated discovery, or the emergence, of order parameters of the system.

In this article, we demonstrate that the AaaA principle works for classical spin systems, by showing that through the AaaA principle the known order parameters emerge. We present three examples: (i) two-dimensional ferromagnetic and antiferromagnetic Ising models, (ii) a two-dimensional spin model with competing stripe orders, and (iii) the fully connected Sherrington--Kirkpatrick (SK) spin-glass model. The first model (i) is the most basic among physics, and the emergent order parameter is the magnetization which is a coordinated average of spins. The second example offers a check if both of the two competing orders in the model can emerge. The last example is about the emergence of the replica overlap as a collective predictive order. In all of these examples, the collective predictive order, which automatically emerges from the spin observation data only, is shown to agree closely with the known order parameter,
up to a nonlinear map.

What needs to be emphasized as for the AaaA principle is that for the emergence of the order parameter one does not need to know the structure of the Hamiltonian. Only the observed spin configurations are necessary to pin down the order parameter. Readers know that in physics textbooks normally the order parameters are introduced with the help of symmetries of the Hamiltonian, since spontaneous symmetry breaking caused by the condensation of the order parameter determines the phase diagram of the system. On the other hand, our AaaA principle is completely data-driven; the order parameters emerge without any information on microscopic structure of the Hamiltonian, such as symmetries. 

{
Mathematically, our CPC formulation belongs to
standard latent-variable modeling: its learning objective is the
negative of the evidence lower bound (ELBO), with an additional
parameter regularization term~\cite{Bishop:2006PRML,Neal:1998EM}.
Our emphasis is on an architecture designed to make the collective
description interpretable, with a shared variable connecting the
local probabilistic prediction models assigned to individual sites, and on the decentralized communication structure to minimize the free energy.
We seek collective predictive order as a compact description that
agents use to predict their own observations and that helps us
understand collective behavior.
The present examples examine its correspondence with familiar
physical order parameters.
The general relations among predictive usefulness, human
interpretability, thermodynamic order, and phase transitions remain open.
}

In this paper, we examine concretely the AaaA principle in two stages:
we first identify collective variables and their thermodynamic properties using centralized CPC (Secs.~\ref{sec:emergent-order} and \ref{sec:emergent-potential}), and then investigate how an ordering pattern can be learned through communication between individual sites (Sec.~\ref{sec:agentic-communication}).

In Sec.~\ref{sec:emergent-order}, we test both the correspondence of the learned coordinates with known order parameters and their ability to predict unobserved spins from observations at other sites.
In Sec.~\ref{sec:emergent-potential}, we use the equilibrium distribution of the learned coordinate to reconstruct a Landau-like free-energy potential for the ferromagnetic and antiferromagnetic Ising models.
The resulting potentials evolve from two wells to one on heating, and yield finite-size estimates of the transition temperature consistent with the known critical point.
Thus, the construction connects the discovery of a collective variable to the thermodynamics of its fluctuations.

In Sec.~\ref{sec:agentic-communication}, each site acts as an agent:
it proposes a value of the shared variable using its own observed spin, while neighboring sites assess that proposal using their local probabilistic prediction models. 
Alternating these exchanges with updates of the local probabilistic prediction models learns a common ferromagnetic pattern from random initial weights.
This procedure is called Metropolis-Hastings naming game (MHNG) \cite{Taniguchi:2023MHNG,Inukai:2023RMHNG} in the context of CPC, and we apply a generalized form of MHNG to the spin system, to find out that the agentic communication in fact properly learns the order parameter.
This provides a concrete realization of the AaaA principle:
the microscopic agents learn how their individual observations relate to a shared macroscopic description.

We believe that this novel approach to emergence of order parameters or macroscopic variables opens a door toward revealing the mystery of emergence. From the viewpoint of a technique for solving physics inversely, in Sec.~\ref{sec:conclusion} we compare our AaaA principle with other machine learning methods for automated discovery of order parameter and phase transition.


\section{Collective predictive coding}
\label{sec:cpc}

An order parameter captures the behavior of many spins in a common macroscopic
variable. CPC explores a similar idea for language: shared words help individuals
with different observations describe and predict the same world
\cite{Taniguchi:2024CPC,Taniguchi:2025CPCMS}. Computer models show that
communication can produce common names for objects and help agents learn object
categories \cite{Taniguchi:2023MHNG,Inukai:2023RMHNG}. 

Predictive coding (PC) describes how an agent predicts sensory
observations and updates its internal representation of their causes using prediction errors~\cite{Friston:2005PC}.
Friston's variational formulation relates such inference and learning to the minimization of an information-theoretic free energy~\cite{Friston:2010FEP}, of the form introduced below in Eq.~\eqref{eq:cpc-single-F}.
\delete{CPC extends this perspective to multiple agents, whose different observations are linked by a shared variable~\cite{Taniguchi:2024CPC}.}\add{CPC extends this perspective to multiple agents to model symbol emergence systems, including human society, whose members' different observations are linked by a shared variable \cite{Taniguchi:2024CPC,Taniguchi:2026SES}. Each agent predicts its own observations, while the collective system integrates their distributed observations through this variable.}\comment{CPC-05: State the SES motivation of the extension from individual predictive coding.}

For example, suppose agents are humans, and all humans look at a single apple in reality. Depending on from which angle each human observes the apple, the picture of the apple may differ for each human observation. \delete{Then humans talk to each other to identify what they are looking at. This is the situation of shared attention. Finally, after the communication, humans arrive at a concept or a word ``apple" which can predicts the situation well and with which any attending human can predict what other humans interpret their observation. CPC is to quantify this situation through a generative model and information free energy minimization.}\add{Then humans communicate to establish a shared symbol (a sign), such as the word ``apple,'' for the object they are looking at. This is a situation of joint attention. Through communication and perceptual learning, they develop a shared use of the sign, grounded in each individual's perceptual categories. Each person uses the sign to predict their own observations, while the sign links these different perspectives into a collective description of the object. CPC quantifies this situation through a generative model and information-theoretic free-energy minimization.}\comment{CPC-06: Make the shared sign explicit, use joint attention, and distinguish shared sign use from identical internal categories.}

Following
Sec.~\ref{sec:introduction}, we treat atoms as agents and seek a shared
variable that predicts their spin configurations: a candidate order parameter
learned from observations alone, without using the Hamiltonian.
We now formulate this idea as a variational problem, specifying in turn the
data, the probability distributions, and the quantities to be varied.

\subsection{Local observations and a shared variable}
\label{sec:cpc-observations}

Let $i=1,\ldots,N$ label agents and $a=1,\ldots,M$ label observed configurations.
The fixed input is the data set
\begin{equation}
 \mathcal D=\{S^{(a)}\}_{a=1}^{M},\qquad
 S^{(a)}=(s_1^{(a)},\ldots,s_N^{(a)}),
 \label{eq:cpc-data}
\end{equation}
where $s_i^{(a)}$ is the local observation of agent $i$ in configuration $a$.
For a spin system, $i$ labels lattice sites and $s_i^{(a)}=\pm1$.
All observations are held fixed throughout the variation which we describe below.

We introduce a shared, unobserved (\emph{latent}) variable $w\in\mathbb R^K$.
In applications of CPC to symbol emergence, this variable represents a common
name for an object observed by different agents. 
Here we allow a continuous
$w$ as a candidate coordinate of the collective spin state. We relate it to
the observations through the joint probability distribution
\begin{equation}
 p_{\alpha,\theta}(S,w)
 =p_\alpha(w)\prod_{i=1}^{N}p_{\theta^{(i)}}^{(i)}(s_i\mid w).
 \label{eq:cpc-model}
\end{equation}
Equation~\eqref{eq:cpc-model} is called a \emph{generative model}: first draw
$w$ from the normalized distribution $p_\alpha(w)$, then draw each $s_i$ from
$p_{\theta^{(i)}}^{(i)}(s_i\mid w)$. The result is a model configuration $S$.
``Generation'' means this statistical sampling procedure, rather than a
physical evolution of the spins.
\add{Each factor $p_{\theta^{(i)}}^{(i)}(s_i\mid w)$ describes an agent's prediction of its own observation, whereas their product describes the collective prediction conditioned on $w$. The agents connected through $w$ can thus be viewed as an integrated collective cognitive system, or a ``super-agent'' \cite{Taniguchi:2025System0123}. Here $w$ provides shared conditioning information for the local predictions, with its prior distribution specified by $p_\alpha(w)$.}\comment{CPC-07: Add the collective cognitive-system perspective (System 0/1/2/3, Sec. 3.2); super-agent summarizes that perspective. The shared variable is not the prior distribution itself.} 

The product in Eq.~\eqref{eq:cpc-model} assumes that the observations are
conditionally independent given $w$; correlations between agents are
represented through this common variable. Within each configuration all
agents refer to the same latent state, while different configurations have
their own states. The parameters
$\alpha$ and $\theta=(\theta^{(1)},\ldots,\theta^{(N)})$ are common to the entire
data set. The latent dimension $K$ and the families of normalized distributions
$\{ p_\alpha(w), p_{\theta^{(i)}}^{(i)}(s_i\mid w) \}$ are modeling choices, 
analogous in physics research to choosing a class of actions before varying its fields.
The theoretical heart of CPC is this starting assumption of the structure \eqref{eq:cpc-model}.

It should be noted that, the agents' inference task proceeds in the
direction reverse to the generation described above: combining their local observations, they try to determine \delete{a distribution over the common $w$, which is the learning process that we describe later.}\add{a distribution over the common $w$. This inference alternates with learning the parameters of the local probabilistic prediction models, as described below.}\comment{CPC-08: Distinguish latent-state inference from parameter learning, consistently with the following subsections.} The measured configurations themselves remain fixed during the learning. The learning determines the parameters $\alpha$ and $\theta=(\theta^{(1)},\ldots,\theta^{(N)})$.

\subsection{Variational principle}
\label{sec:cpc-variational}

\subsubsection{Variational free energy}
\label{sec:cpc-variational-energy}

CPC formulates the learning process as a variational principle.
For each configuration labeled by $a$, introduce an auxiliary probability density $q_a(w)$, subject to $q_a(w)\geq0$ and $\int dw\,q_a(w)=1$.
The variational degrees of freedom are the functions $\{q_a\}$ and the common
parameters $\alpha,\theta$; $w$ itself is the integration variable.
Define
\begin{align}
 \mathcal F[\{q_a\},\alpha,\theta\mid\mathcal D]
 & \equiv \sum_{a=1}^{M}F_a+\Omega_{\mathrm{reg}}(\alpha,\theta),
 \label{eq:cpc-total-F}\\
 F_a
 &\equiv \int dw\,q_a(w)
 \log\frac{q_a(w)}{p_\alpha(w)\prod_i
 p_{\theta^{(i)}}^{(i)}(s_i^{(a)}\mid w)}.
 \label{eq:cpc-single-F}
\end{align}
Each observed value $s_i^{(a)}$ is inserted into its corresponding local likelihood,
averaged with the distribution $q_a$ for that configuration, and finally summed
over all configurations. The optional regularization $\Omega_{\mathrm{reg}}$ constrains the model parameters; the basic formulation has $\Omega_{\mathrm{reg}}=0$.


The reason we define the free energy like this will be clarified later, but here we note that
one can further decompose the free energy and look at the decomposed meaning. 
Consider the decomposition
\begin{equation}
 F_a=D_{\rm KL}(q_a\Vert p_\alpha)
 -\sum_i\bigl\langle
 \log p_{\theta^{(i)}}^{(i)}(s_i^{(a)}\mid w)
 \bigr\rangle_{q_a},
 \label{eq:cpc-decomposition}
\end{equation}
where $\langle f\rangle_q\equiv \int dw\,q(w)f(w)$.
Here $D_{\rm KL}(q\Vert p)\equiv\int dw\,q(w)\log[q(w)/p(w)]$
is the Kullback--Leibler divergence.
The first term measures the difference between $q_a(w)$ and $p_\alpha(w)$;
the second measures the loss in predicting the observations. 

The variational principle is
\begin{equation}
 (\{q_a^*\},\alpha^*,\theta^*)
 \in\operatorname*{arg\,min}_{\{q_a\},\alpha,\theta}
 \mathcal F[\{q_a\},\alpha,\theta\mid\mathcal D]
 \label{eq:cpc-variational-principle}
\end{equation}
with $q_a(w)\geq0$ and $\int dw\,q_a(w)=1$ for each $a$, while $\mathcal D$
is held fixed. Thus both the configuration-dependent distributions $q_a(w)$
and the parameters $\alpha,\theta$ shared by all configurations are varied.
This free energy is a dimensionless information quantity.
The physical Hamiltonian and temperature enter only through the
observed configurations ${\cal D}$ indirectly; evaluating $\mathcal F$ requires the data and the chosen generative model.
The meaning of the variational principle \eqref{eq:cpc-variational-principle} will be made clear as we explain the inference and the learning, below.

\subsubsection{Inference of latent states}
\label{sec:cpc-inference}

First hold $\alpha,\theta$ fixed and vary $q_a$.
The stationary condition with a normalization multiplier $\xi_a$ is
$\log q_a+1-\log p_{\alpha,\theta}(S^{(a)},w)+\xi_a=0$, giving
\begin{equation}
 q_a^*(w)=
 \frac{p_\alpha(w)\prod_i
 p_{\theta^{(i)}}^{(i)}(s_i^{(a)}\mid w)}
 {p_{\alpha,\theta}(S^{(a)})}
 =p_{\alpha,\theta}(w\mid S^{(a)}),
 \label{eq:cpc-posterior}
\end{equation}
where $p_{\alpha,\theta}(S) \equiv\int dw\,p_{\alpha,\theta}(S,w)$ solves $\xi_a$ to give the first equality, and the last equality in Eq.~\eqref{eq:cpc-posterior} uses Bayes' theorem:
normalizing the joint distribution at the observed $S^{(a)}$ gives the
conditional distribution of $w$.
Thus the optimized $q_a^*(w)$ is the conditional distribution
$p_{\alpha,\theta}(w\mid S^{(a)})$ determined by the fixed observation
$S^{(a)}$ through Eq.~\eqref{eq:cpc-posterior}.
This step is called \emph{inference}.\footnote{In Bayesian terminology,
$p_\alpha(w)$ and $p_{\alpha,\theta}(w\mid S^{(a)})$ are called the
``prior'' and ``posterior'' distributions, respectively. These terms distinguish
distributions without and with conditioning on a particular configuration
$S^{(a)}$. In the present variational problem, all observations are fixed from
the outset, and $p_\alpha(w)$ can itself be learned from the full data set.}
Indeed, the identity
\begin{equation}
 F_a=D_{\rm KL}\!\left(q_a\Vert
 p_{\alpha,\theta}(w\mid S^{(a)})\right)
 -\log p_{\alpha,\theta}(S^{(a)})
 \label{eq:cpc-posterior-KL}
\end{equation}
shows that $q_a(w)=q_a^*(w)$ of Eq.~\eqref{eq:cpc-posterior} minimizes $F_a$,
since $D_{\rm KL}\geq0$.
Restricting $q_a$ to a chosen family gives an approximate inference scheme.

This clarifies the reason of our definition of the free energy by introducing auxiliary distribution $q_a$.
It is a strange way of writing, but the point is the structure of $p$ given in \eqref{eq:cpc-model}: we want to determine the parameters under the constraint of the structure of the generative model.

\subsubsection{Learning the configuration distribution}
\label{sec:cpc-learning}

Hold $\{q_a\}$ fixed and minimize $\mathcal F$ over $\alpha,\theta$.
This is \emph{learning}: all configurations contribute to a common predictive
rule. For example, each local probabilistic prediction model has gradient
\begin{equation}
 \frac{\partial\mathcal F}{\partial\theta^{(i)}}
 =-\sum_a\left\langle
 \frac{\partial}{\partial\theta^{(i)}}
 \log p_{\theta^{(i)}}^{(i)}(s_i^{(a)}\mid w)
 \right\rangle_{q_a}
 +\frac{\partial\Omega_{\mathrm{reg}}}{\partial\theta^{(i)}}.
 \label{eq:cpc-learning-gradient}
\end{equation}
Why is this gradient descent actually the learning process with use of the data ${\cal D}$? To answer this question, let us look at the free energy which is minimized with $q_a^*$:
\begin{equation}
 \mathcal J(\alpha,\theta)\equiv\min_{\{q_a\}}\mathcal F
 =-\sum_a\log p_{\alpha,\theta}(S^{(a)})+\Omega_{\mathrm{reg}}(\alpha,\theta).
 \label{eq:cpc-marginal-objective}
\end{equation}
The first term in fact measures the deviation of the model prediction against the data ${\cal D}$, as follows. To explain this in an easier way, let us define
\begin{equation}
 \widehat p_{\mathcal D}(S)
 =\frac{1}{M}\sum_{a=1}^{M}\mathbf 1\{S=S^{(a)}\},
 \label{eq:cpc-empirical-distribution}
\end{equation}
where $\mathbf 1$ is one when the condition holds and zero otherwise.
This is called an empirical distribution.
Equation~\eqref{eq:cpc-marginal-objective} then becomes
\begin{equation}
 \frac{\mathcal J(\alpha,\theta)}{M}
 =
 D_{\rm KL}(\widehat p_{\mathcal D}\Vert p_{\alpha,\theta})
 +H(\widehat p_{\mathcal D})
 +\frac{\Omega_{\mathrm{reg}}(\alpha,\theta)}{M},
 \label{eq:cpc-empirical-KL}
\end{equation}
with the entropy of the empirical distribution $H(\widehat p_{\mathcal D}) \equiv -\sum_S\widehat p_{\mathcal D}(S) \log\widehat p_{\mathcal D}(S)$.
Thus, the first term tells us that minimizing $\mathcal J$ brings the model's configuration distribution $p_{\alpha,\theta}(S)$ closer to the observed frequency distribution $\widehat p_{\mathcal D}(S)$ in KL divergence. Here we have ignored the regularization $\Omega_{\mathrm{reg}}=0$ for simplicity, and the entropy of the empirical distribution $H(\widehat p_{\mathcal D})$ is of no importance because it is a fixed number given by the data.

This is how minimizing the variational free energy incorporates the statistics
of the observations into the model. Regularization balances this fit against
the chosen constraints on the parameters.

\subsubsection{Numerical strategy: Alternating minimization and collective prediction}
\label{sec:cpc-optimization}

Inference and learning are called respectively the E-step and M-step of
expectation--maximization (EM). The variational principle
\eqref{eq:cpc-variational-principle} is implemented numerically by alternating these steps
as follows:
\begin{enumerate}
 \item Fix $\mathcal D$, the model families, and $\Omega_{\mathrm{reg}}$;
 initialize $\alpha,\theta$.
 \item At fixed $\alpha,\theta$, determine each $q_a(w)$ using
 Eq.~\eqref{eq:cpc-posterior} (E-step).
 \item At fixed $\{q_a\}$, update the common parameters $\alpha,\theta$
 to decrease $\mathcal F$ (M-step).
 \item Repeat the two updates, monitoring $\mathcal F$ and the parameters
 for convergence.
\end{enumerate}
These are two alternating minimizations of the same functional.

Two notes on the minimization: first, 
different initializations may lead to different local minima and should be
compared. Second, with approximate inference, the corresponding variational bound
will be optimized (note that the equality in Eq.~\eqref{eq:cpc-marginal-objective} requires unrestricted optimization over $\{q_a\}$ so it may not be achieved during the learning).

\subsection{Extracting the latent constituents}
\label{sec:cpc-latent-extr}

After the inference and the learning loops are numerically performed, the model parameters are determined. Then, from this determined generative model, how can we see the relation between the latent $w$ and the data $S$? This reading-off method from the generative model leads to the automated discovery of a macroscopic variable made of microscopic variables.

Let us take a new configuration $S$. 
After learning, parameters $\alpha^*,\theta^*$ determine the conditional
distribution $p_{\alpha^*,\theta^*}(w\mid S)$ for the new configuration $S$
through Eq.~\eqref{eq:cpc-posterior}. A macroscopic coordinate can be extracted as
\begin{equation}
 \widehat w(S)=\int dw\,w\,p_{\alpha^*,\theta^*}(w\mid S).
 \label{eq:cpc-coordinate}
\end{equation}
This is the relation between the latent variable $w$ and the configuration $S$.
The collective aspect of this expression is explicitly seen in Eq.~\eqref{eq:cpc-posterior}: each agent contributes $\log p_{\theta^{(i)}}^{(i)}(s_i^{(a)}\mid w)$ to $\log q_a^*(w)$, together with the common term $\log p_\alpha(w)$ and a normalization constant. The choice of $w$ is learned so that the whole set of agents' inference agree with each other as a single $w$, thus the architecture is optimized so that the macroscopic variable $w$ represents the configuration $S$ appropriately.


The learned coordinate provides a candidate order parameter whose physical
meaning is established by its predictive power. Its physical significance is measured by its relation to symmetries and phase transitions. Its form depends on the chosen model class, while its parameters are determined from the observed configurations. 
We call the learned coordinate $\widehat w(S)$ the {\it collective predictive order}.
In the following sections we examine this correspondence for classical spin systems.


\section{Emergent order parameter}
\label{sec:emergent-order}

We now ask whether the collective predictive order defined in Sec.~\ref{sec:cpc} numerically emerges and acquires the
meaning of a physical order parameter. Three examples test progressively
different requirements: (i) a scalar magnetization, (ii) two competing stripe orders,
and (iii) an overlap between replicas of the fully connected SK spin glass. Here (i) is for demonstrating the reproduction of the most popular spin model, while (ii) is a study whether our CPC can actually count a necessary number of order parameters. And (iii) is of course to study whether the CPC can find one of the most nontrivial order parameters known, the replica order. In all the examples, as we have emphasized, the Hamiltonians below specify only the ensembles used to generate the data; they are not supplied to CPC.

\subsection{A local probabilistic prediction model and the learned coordinate}
\label{sec:s3-local-model}

For Ising observations $s_i=\pm1$, take $w\in\mathbb R^K$ and specialize
Eq.~\eqref{eq:cpc-model} to the following generative model
\begin{equation}
 \begin{split}
 p_0(w)&=(2\pi)^{-K/2}e^{-|w|^2/2},\\
 p_{\theta^{(i)}}^{(i)}(s_i\mid w)
 &=\frac{e^{s_i(b_i+W_i\cdot w)}}{2\cosh(b_i+W_i\cdot w)}.
 \end{split}
 \label{eq:s3-local-model}
\end{equation}
This model realizes its conditional mean as $\langle s_i\rangle_w=\tanh(b_i+W_i\cdot w)$. 
Each site has its own freely learned $b_i$ and $K$-component $W_i$, and the model parameters, previously written as $\{\alpha, \theta_i\}$ are now $\{b_i, W_i\}$ (and $K$).
No symmetry such as uniform, staggered, or stripe pattern is imposed on the training parameters. The regularization is
\begin{equation}
 \Omega_{\mathrm{reg}}
 =\frac{\lambda_W}{2}\sum_i|W_i|^2
 +\frac{\lambda_b}{2}\sum_i b_i^2,
 \qquad \lambda_b=\max(1,0.1\lambda_W).
 \label{eq:s3-regularization}
\end{equation}
The objective uses the sum over configurations in
Eq.~\eqref{eq:cpc-total-F}, without dividing it by the sample count.
At fixed parameters, Eq.~\eqref{eq:cpc-posterior} determines the conditional
distribution of $w$ by Bayes' theorem. We evaluate its integrals by numerical
quadrature and update the site parameters by Newton steps on the concave
local learning objectives. Thus this implementation uses explicit probability
distributions, without a neural network. 

We use the conditional mean
$\widehat w(S)$ in Eq.~\eqref{eq:cpc-coordinate} as the candidate order
parameter itself --- the collective predictive order.
In our model architecture \eqref{eq:s3-local-model}, 
the relation between the collective predictive order $\widehat w(S)$ and the familiar spin averages can be made explicit. For the case $K=1$, define a particular combination of the spins
\begin{equation}
 h_W(S)\equiv \sum_iW_i s_i.
 \end{equation}
 Then, with the following definition of an unnormalized weight
 \begin{equation}
 \rho(w) \equiv \exp\!\left[-\frac{w^2}{2}
 -\sum_i\log 2\cosh(b_i+W_iw)\right], 
 \label{eq:s3-projection}
\end{equation}
we find
\begin{equation}
 \widehat w(S)=G(h_W(S)),\qquad
 G(h)=\frac{\int dw\,w\rho(w)e^{wh}}{\int dw\,\rho(w)e^{wh}},
 \qquad G'(h)=\operatorname{Var}_{h}(w)>0.
 \label{eq:s3-nonlinear-coordinate}
\end{equation}
To see why only $h_W(S)$ appears here, write the joint distribution as
\begin{equation*}
 p(S,w)=(2\pi)^{-1/2}e^{\sum_i b_i s_i}\rho(w)e^{w h_W(S)}.
\end{equation*}
When inferring $w$, the observed spins $S$ are held fixed. Hence
$e^{\sum_i b_i s_i}$ is the same factor for every possible $w$: it cannot
favor one value of $w$ over another. Dividing by $p(S)=\int dw\,p(S,w)$
to obtain $p(w\mid S)$ cancels this factor. This is analogous to the
cancellation of a constant energy shift in a normalized Boltzmann
probability. The $b_i$ still affect $\rho(w)$ and therefore the map $G$.

The physical point is that, for nearly uniform or staggered weights,
the discovered collective predictive order $\widehat w(S)$ is
approximately related to the corresponding magnetization through
the nonlinear monotonic map in Eq.~\eqref{eq:s3-nonlinear-coordinate}. 
Eq \eqref{eq:s3-nonlinear-coordinate} shows that once the parameters have been learned, the inferred coordinate depends on the many observed spins only through one weighted sum, $h_W(S)$. Two configurations with the same $h_W$ give the same $\widehat w$, and the monotonicity $G'(h)>0$ means that $G$ preserves the ordering of these weighted sums. 

If the learned $W_i$ are uniform, this sum is proportional to the magnetization; if their signs alternate between sublattices with equal magnitude, it is proportional to the staggered magnetization. 
Thus the weights $W_i$ select which collective spin pattern is represented, while $G$ converts its amplitude into the collective predictive order. Whether the
selected pattern is a physical order parameter is tested below. The overall sign is arbitrary; for $K>1$, rotations of the latent axes are also arbitrary. Their scale is fixed by the chosen model convention, not by a thermodynamic normalization.

\subsection{Two criteria for a successful discovery}
\label{sec:s3-prediction}

We judge the learned coordinate by two criteria. First, does it reproduce a
known physical order parameter, allowing for the change of coordinates
described above? Second, does knowing this coordinate help predict spins
that have not been observed? The first criterion identifies the physical
meaning of the coordinate; the second tests whether it captures information
shared among sites. 
{
We use both checks in the present examples: even in a disordered
phase, a coordinate can track magnetization fluctuations without
appreciably helping to predict other spins.
}

To test these criteria on data not used to determine the model, we divide
the available configurations into three sets. The \emph{training set} is
used to learn $b_i$ and $W_i$. The \emph{validation set} is used to choose
among fits obtained with different regularization strengths and, where
compared, different latent dimensions $K$. The \emph{test set} is used only
for the final assessment, after these choices and all parameters have been
fixed. The precise splits are given for each example below. 
Our training uses data with different several temperatures, 
but supplies neither temperature labels nor coupling constants, lattice coordinates, or known order parameters. 
Only the site labels identify which observations belong to the same site across configurations.

For the model selection (such as a proper choice of $K$), we use the validation data set. 
We ask how much probability a learned model assigns to
each validation configuration $S$. Its negative log likelihood
(NLL) is $-\log p_{\alpha,\theta}(S)$, where the probability is obtained by
integrating over $w$. A smaller NLL means that the observed configuration
is assigned a larger probability, meaning that, for the validation data $S$, a smaller NLL is with a better model. We compare the mean NLL per site,
\begin{equation*}
 -\frac{1}{N|\mathcal V|}\sum_{S\in\mathcal V}
 \log p_{\alpha,\theta}(S),
\end{equation*}
where $\mathcal V$ is the validation set, and select the model architecture with the lowest value of this mean NLL. This comparison determines the model to be further tested with the two criteria described.

We assess the first criterion by comparing the coordinate with the known
order parameter on test configurations. For the second criterion, we shall perform the following prediction task on the validated model: We show to the model only a
randomly chosen set $O$ containing $25\%$ of the sites in each test
configuration, and hide the remaining $75\%$. Using only the revealed spins
$S_O$, we infer a distribution of $w$ and predict a hidden spin $s_j$:
\begin{equation}
 \begin{split}
 q_O(w)&=\frac{p_0(w)\prod_{i\in O}p_{\theta^{(i)}}^{(i)}(s_i\mid w)}
 {\int dw'\,p_0(w')\prod_{i\in O}p_{\theta^{(i)}}^{(i)}(s_i\mid w')},\\
 p_{\mathrm{pred}}(s_j\mid S_O)
 &=\int dw\,q_O(w)p_{\theta^{(j)}}^{(j)}(s_j\mid w),\quad j\notin O.
 \end{split}
 \label{eq:s3-heldout}
\end{equation}
We then reveal the hidden spins and evaluate the probability assigned to
their actual values. We define the improvement of the NLL as
\begin{equation}
 \Delta\mathrm{NLL}
 =\left\langle\log p_{\mathrm{pred}}(s_j\mid S_O)
              -\log p_{\mathrm{ind},j}(s_j)\right\rangle_{S,O,j\notin O},
 \label{eq:s3-predictive-gain}
\end{equation}
where the average is over test configurations, random choices of observed
sites, and hidden sites. 
The second term which we introduced as a reference value is made by an independent-site prediction
$p_{\mathrm{ind},j}(s_j)$ which is defined by using only the frequency of that value at site $j$
in the training data. For example, if spin $j$ is up in half of the training
configurations, this reference assigns probability $1/2$ to up, regardless
of the other observed spins. It therefore uses no information shared
between sites.

Thus, this \eqref{eq:s3-predictive-gain} is the independent-site prediction NLL minus
the CPC prediction NLL: a positive value means that CPC predicts the hidden
spins better. We call this quantity the \emph{prediction gain} below.\footnote{
The gain is dimensionless, and the conventional name for this logarithmic unit is a ``nat.''} The reported gain is per hidden
site. Unlike the NLL for complete configurations used in model selection,
this test measures how much observing some spins helps predict the others.
The predictivity lies at the heart of CPC.


\subsection{Ferromagnetic and antiferromagnetic order}
\label{sec:s3-ferro}

Consider square lattices with periodic boundary conditions and
\begin{equation}
 H_{\mathrm F}=-\sum_{\langle i,j\rangle}s_i s_j,\qquad
 H_{\mathrm{AF}}=+\sum_{\langle i,j\rangle}s_i s_j,
 \label{eq:s3-ising-hamiltonians}
\end{equation}
where the coupling magnitude and Boltzmann constant are set to unity.
For each model, configurations on a $32\times32$ lattice were pooled over
13 temperatures from $1.5$ to $3.5$, whose range includes the known phase transition temperature $T_c=2/\log(1+\sqrt2)$. At each temperature, 240 configurations were used for
learning, 60 for validation, and 200 for testing. We used $K=1$ and 101
quadrature points on $[-4.5,4.5]$ to perform integrations. Among $\lambda_W=1,10,100$, each with three
initializations, validation selected $\lambda_W=1$ for both models. We tried three initializations; one initialization used the leading covariance mode without physical labels\footnote{For the first initialization, we computed the spin covariance
matrix $C_{ij}=\langle s_i s_j\rangle-\langle s_i\rangle\langle s_j\rangle$
from the training configurations and initialized the weights $W_i$
along its eigenvector with the largest eigenvalue. This direction
captures the largest collective fluctuation in the data; no phase
labels or known order parameters were supplied.}, while the other two were random.

\begin{figure}[t]
 \centering
 \includegraphics[width=0.98\textwidth]{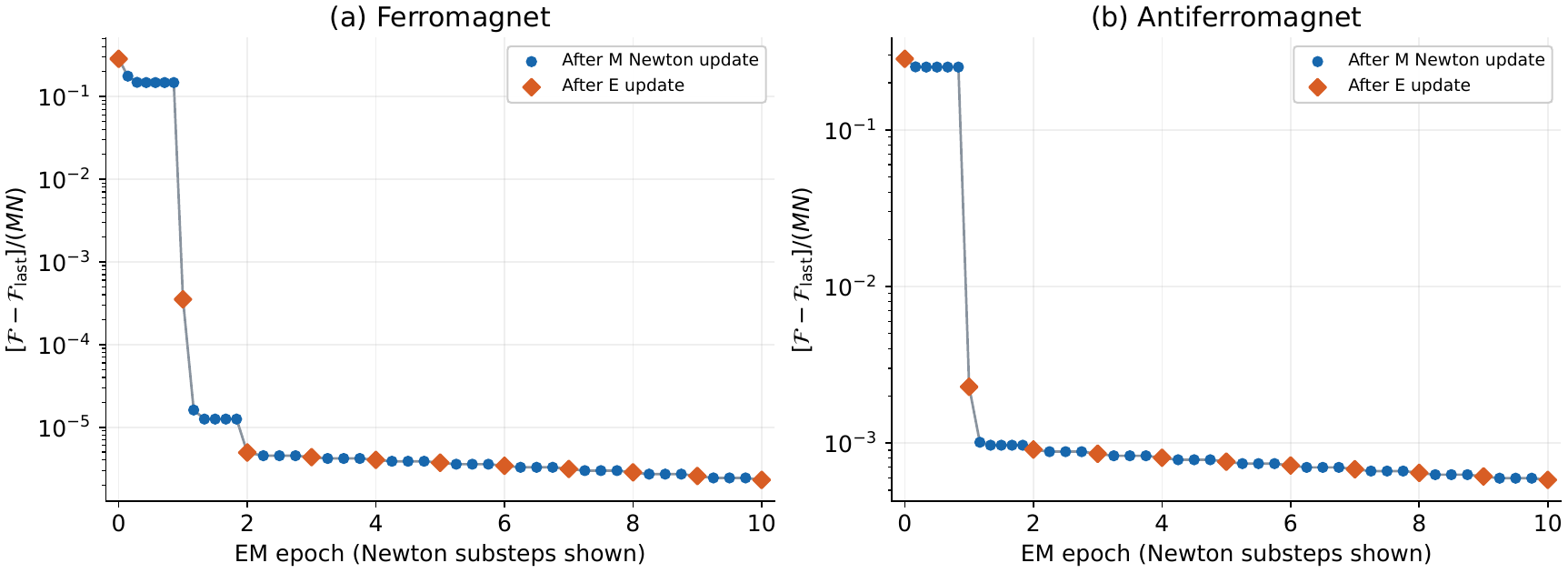}
 \caption{E and M updates of the regularized free energy for the selected
 ferromagnetic (left) and antiferromagnetic (right) cases. The first ten
 epochs are shown. Blue circles mark individual Newton updates within an
 M-step; orange diamonds mark the subsequent E-step, with epoch zero
 denoting the initial E-step. Intermediate Newton updates are spaced
 evenly within each epoch for display; the horizontal axis is not elapsed
 time. The vertical axis is the difference from the final recorded free
 energy, divided by the number of training configurations and sites, on
 a logarithmic scale. $\mathcal F_{\mathrm{last}}$ is the value at numerical
 termination, not an assumed exact minimum.}
 \label{fig:s3-ising-em-descent}
\end{figure}

Figure~\ref{fig:s3-ising-em-descent} resolves the decrease of the regularized
free energy into its E and M updates for the selected fits, both of which
started from random weights. At fixed parameters, the E-step updates all
$q_a(w)$ by the quadrature described above. At fixed $\{q_a\}$, the M-step
updates each site's $(b_i,W_i)$ using at most eight Newton iterations,
stopping early when the largest absolute parameter increment is below
$10^{-6}$. Starting after the initial E-step, we count one M-step followed
by an E-step as one epoch. The runs terminated after 20 epochs for the
ferromagnet and 142 for the antiferromagnet, when the relative decrease
of the regularized free energy between successive epochs fell below
$5\times10^{-7}$. The final values were
$\mathcal F_{\mathrm{last}}/(MN)=0.407869$ and $0.409239$, respectively,
with $M=3120$ training configurations and $N=1024$ sites. To resolve the
small decreases following the initial rapid drop, the figure shows the
difference from the final value on a logarithmic scale. Evaluating
$\mathcal F$ after each update, while holding $q_a$ fixed during the M-step,
confirms that both E and M updates lower it, up to numerical roundoff.

\begin{figure}[t]
 \centering
 \includegraphics[width=0.84\textwidth]{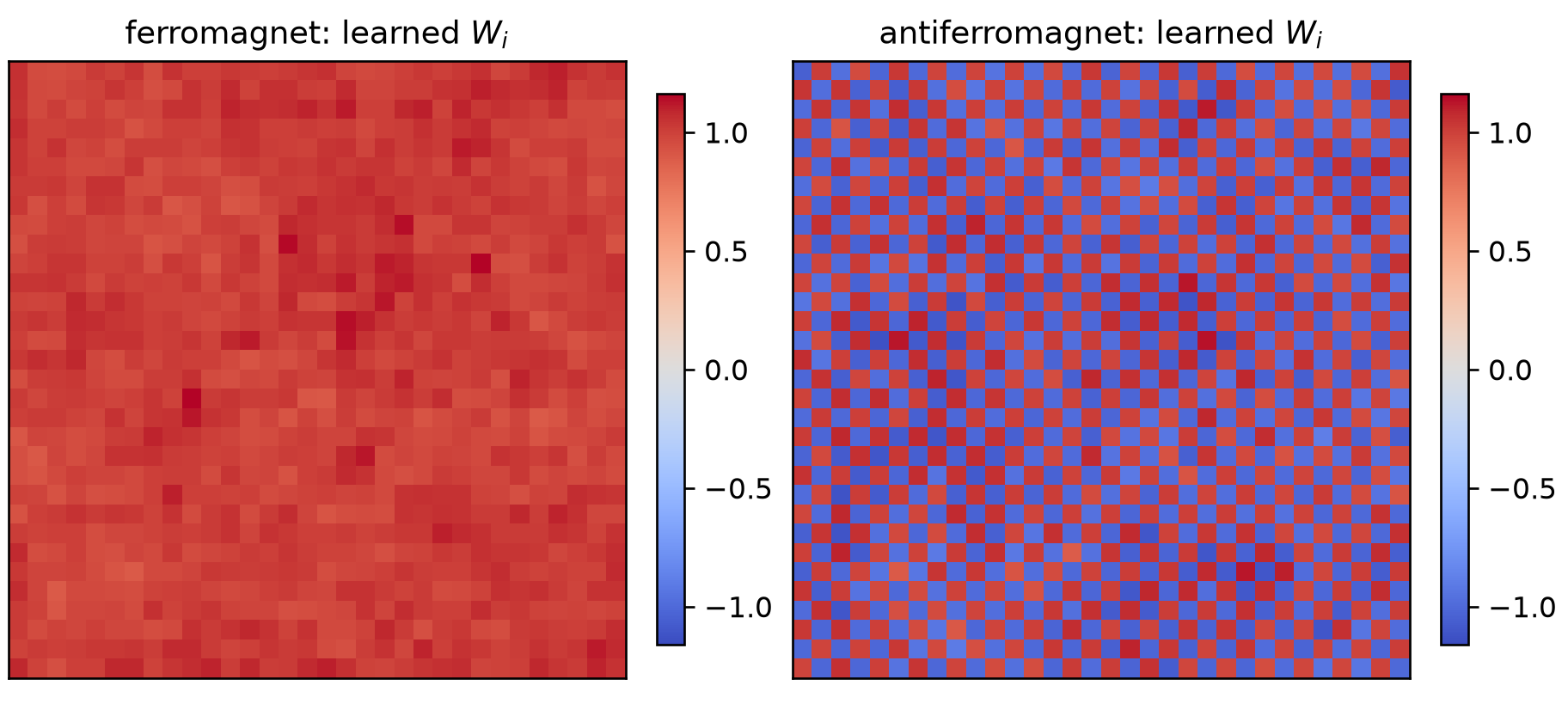}
 \caption{Spatial weights $W_i$ learned independently from ferromagnetic
 (left) and antiferromagnetic (right) configurations. The uniform and
 staggered patterns are results of learning, rather than constraints.}
 \label{fig:s3-ising-modes}
\end{figure}

Figure~\ref{fig:s3-ising-modes} displays the learned weights. Their normalized
absolute overlaps with the uniform and staggered vectors are respectively
$0.99913$ for the ferromagnet and $0.99916$ for the antiferromagnet. Overlaps
with the incorrect pattern are below $2.5\times10^{-4}$. Hence the same
learning rule selects either
\begin{equation}
 m_{\mathrm F}(S)=\frac1N\sum_i s_i,
 \qquad
 m_{\mathrm{AF}}(S)=\frac1N\sum_i(-1)^{x_i+y_i}s_i
 \label{eq:s3-magnetizations}
\end{equation}
as the relevant spatial projection, although neither expression was supplied.

\begin{figure}[t]
 \centering
 \includegraphics[width=0.94\textwidth]{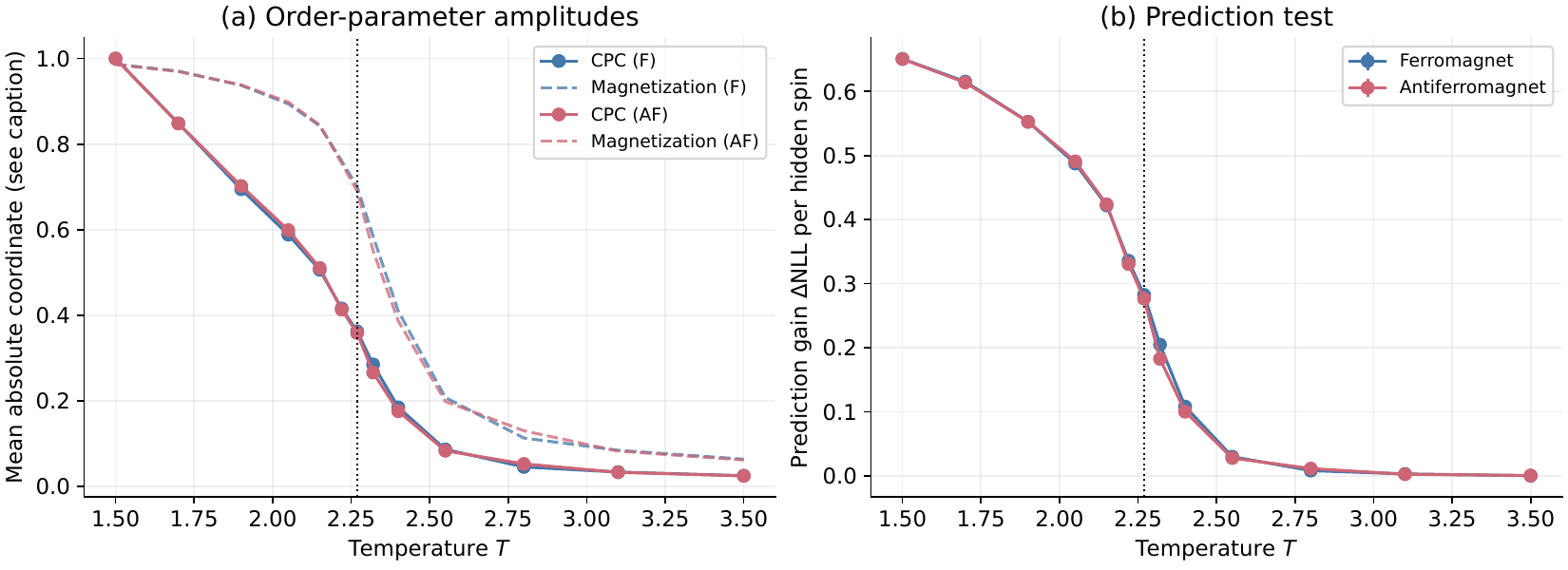}
 \caption{Temperature dependence and predictive performance of the learned
 Ising coordinate. (a) Solid curves: mean absolute collective predictive order divided
 by its value at $T=1.5$. Dashed curves: mean absolute magnetization for
 the ferromagnet (F) and staggered magnetization for the antiferromagnet
 (AF), without rescaling. The nonlinear maps in
 Eq.~\eqref{eq:s3-quintic-maps} relate the learned coordinate to these
 conventional order parameters. (b) Prediction gain $\Delta\mathrm{NLL}$,
 Eq.~\eqref{eq:s3-predictive-gain}, for the hidden $75\%$ of spins given
 the observed $25\%$. Positive values indicate a lower prediction NLL
 than the independent-site reference. 
 Vertical dotted lines mark $T_c$.}
 \label{fig:s3-ising-scan}
\end{figure}

Figure~\ref{fig:s3-ising-scan}(a) compares the temperature dependence of
the learned coordinate with that of the conventional magnetization
($m_{\mathrm F}$ or $m_{\mathrm{AF}}$). The solid curves show
$\langle|\widehat w|\rangle_T/\langle|\widehat w|\rangle_{T=1.5}$, while
the dashed curves show $\langle|m|\rangle_T$. Both\footnote{The normalization factors for the solid curves are $2.47493$ for the ferromagnet and $2.45642$ for the antiferromagnet; the magnetization curves are shown without this rescaling.} decrease across the
transition region around $T_c$.

Their shapes differ: the collective predictive order is a nonlinear function of magnetization, as we prescribed before. 
The nonlinear relation can be estimated explicitly. With the arbitrary sign
chosen so that $\widehat w$ increases with magnetization, the nearly uniform
or staggered weights give $h_W(S)\simeq A_{\mathrm F}m_{\mathrm F}(S)$ or
$A_{\mathrm{AF}}m_{\mathrm{AF}}(S)$, where
$A_{\mathrm F}=\sum_i W_i=1031.98$ and
$A_{\mathrm{AF}}=\sum_i(-1)^{x_i+y_i}W_i=1031.73$, respectively.
Fitting the unnormalized $\widehat w$ against the corresponding $m$ gives
the following odd-quintic approximations to the rescaled maps $G$ in
Eq.~\eqref{eq:s3-nonlinear-coordinate}:
\begin{equation}
 \begin{split}
 G_{\mathrm F}(A_{\mathrm F}m)
 &\simeq 1.47907m-2.49008m^3+3.46003m^5,\\
 G_{\mathrm{AF}}(A_{\mathrm{AF}}m)
 &\simeq 1.40820m-2.23266m^3+3.24455m^5.
 \end{split}
 \label{eq:s3-quintic-maps}
\end{equation}
For each model, the coefficients were fitted using 1300 of the 2600 test
configurations pooled over all temperatures, and evaluated on the remaining
1300, giving $R^2=0.99408$ and $0.99392$, respectively. Thus the different
curve shapes in Fig.~\ref{fig:s3-ising-scan}(a) are consistent with a
nonlinear reparametrization of magnetization. These fits describe the
already learned coordinates; magnetization was not supplied during CPC
training or inference.

We next apply the second criterion of Sec.~\ref{sec:s3-prediction}.
With the learned parameters fixed, we reveal $25\%$ of the sites in each
test configuration and predict the remaining $75\%$ using
Eq.~\eqref{eq:s3-heldout}. Relative to the independent-site prediction based
on training frequencies, the prediction gain $\Delta\mathrm{NLL}$ of
Eq.~\eqref{eq:s3-predictive-gain} is about $0.651$ at $T=1.5$ and $0.28$
at $T_c$ for both models. These positive values mean that observing some
spins improves the prediction of the others. At $T=3.5$ which is higher than the phase transition temperature, the gains are only
$1.5\times10^{-4}$ and $6.1\times10^{-4}$ for the ferromagnet and
antiferromagnet, respectively, in nats per hidden spin. Thus the collective predictive order
still tracks magnetization fluctuations at high temperature, but provides
essentially no predictive improvement over independent sites. 

\subsection{Two competing stripe orders}
\label{sec:s3-stripes}

To test whether more than one order parameter can be identified,
consider the frustrated square-lattice model
\begin{equation}
 H=-J_1\sum_{\langle i,j\rangle}s_i s_j
   +J_2\sum_{\langle\!\langle i,j\rangle\!\rangle}s_i s_j,
 \qquad J_1=J_2=1.
 \label{eq:s3-j1j2}
\end{equation}
The second sum couples diagonal neighbors. At low temperature, vertical
and horizontal stripes compete on the same spins. Their conventional
order parameters, used only in evaluation, are
\begin{equation}
 m_x=\frac1N\sum_{x,y}(-1)^x s_{x,y},\qquad
 m_y=\frac1N\sum_{x,y}(-1)^y s_{x,y}.
 \label{eq:s3-stripe-order}
\end{equation}
Four ordered sectors lie near $(\pm m_0,0)$ and $(0,\pm m_0)$.

We generated $16\times16$ configurations by Metropolis sampling with parallel
tempering at 13 temperatures from $1.2$ to $3.6$, using 140/40/80
configurations per temperature for training/validation/test. After 1400
initial sweeps, samples were saved every ten sweeps. Exact lattice
symmetries---global spin reversal, rotation, and translation---were randomly
applied to reduce sector imbalance. These transformations were not supplied
to CPC. We compared $K=1$ with $K=2$, using respectively 81 points and a
$17\times17$ quadrature grid on $[-4.2,4.2]^K$. Validation selected
$\lambda_W=1$ from $1,10$, with two initializations per choice.

We first ask whether the learned weights reproduce the two stripe
patterns associated with $m_x$ and $m_y$ in
Eq.~\eqref{eq:s3-stripe-order}. These patterns are represented by the
site-dependent vectors $u_{x,i}=(-1)^{x_i}$ and $u_{y,i}=(-1)^{y_i}$.
For $K=2$, we would like to see the the consistency between the two learned weight vectors and the two components of our learned collective predictive order. 
Note that they need not coincide with $u_x$ and $u_y$ individually: they may be linear combinations of the two. Recovering both stripe patterns therefore means that suitable linear combinations of the learned vectors can reproduce
both $u_x$ and $u_y$.

To quantify this, let $W$ be the $N\times K$ matrix whose columns
are the learned weight vectors, and let $U=(u_x,u_y)$.
The matrices
\begin{equation}
 P_W\equiv W(W^{\mathsf T}W)^{-1}W^{\mathsf T},
 \qquad
 P_U \equiv U(U^{\mathsf T}U)^{-1}U^{\mathsf T}
\end{equation}
project a spatial pattern onto the space spanned by the learned
vectors and onto the space spanned by the two stripe patterns,
respectively. We measure their agreement by
\begin{equation}
 \mathcal O \equiv \frac12\operatorname{Tr}(P_WP_U).
 \label{eq:s3-subspace}
\end{equation}
Equivalently, $\mathcal O$ is the average fraction of the squared
length of each stripe vector retained when projected onto the
learned space. It equals one when both stripe patterns can be
fully reconstructed, and zero when the learned vectors are
orthogonal to both patterns.

We obtain $\mathcal O=0.9963$ for $K=2$, showing that the two learned
vectors span almost exactly the space of the two stripe patterns.
For $K=1$, we obtain $\mathcal O=0.4973$, close to the maximum
possible value of $1/2$ for a single learned vector. Thus the $K=1$
model captures one direction within the stripe space, whereas the
two-component model captures both independent stripe directions.

Let us proceed for the second criterion.
There are two distinct probability comparisons. For complete test
configurations, the mean NLL per spin is $0.51631$ for $K=1$ and $0.38433$
for $K=2$, so the two-component model assigns larger probability to the
observed configurations on average. Separately, we perform the masked
prediction test of Sec.~\ref{sec:s3-prediction}: $25\%$ of the spins are
revealed and the remaining $75\%$ are predicted, with all parameters fixed.
At $T=1.2$, the reduction in prediction NLL relative to the same
independent-site reference is $\Delta\mathrm{NLL}=0.6557$ for $K=2$,
compared with $0.4234$ for $K=1$, per hidden spin.
The second component therefore improves both the probability assigned to
complete test configurations and the prediction of their hidden spins. 
This concludes that our CPC can find that spins need at least two components of collective predictive order and $K=2$ is favored.\footnote{
This dimension comparison is within the linear local-field family of Eq.~\eqref{eq:s3-local-model}; it is not a general impossibility result for scalar nonlinear models.}

\begin{figure}[t]
 \centering
 \includegraphics[width=0.60\textwidth]{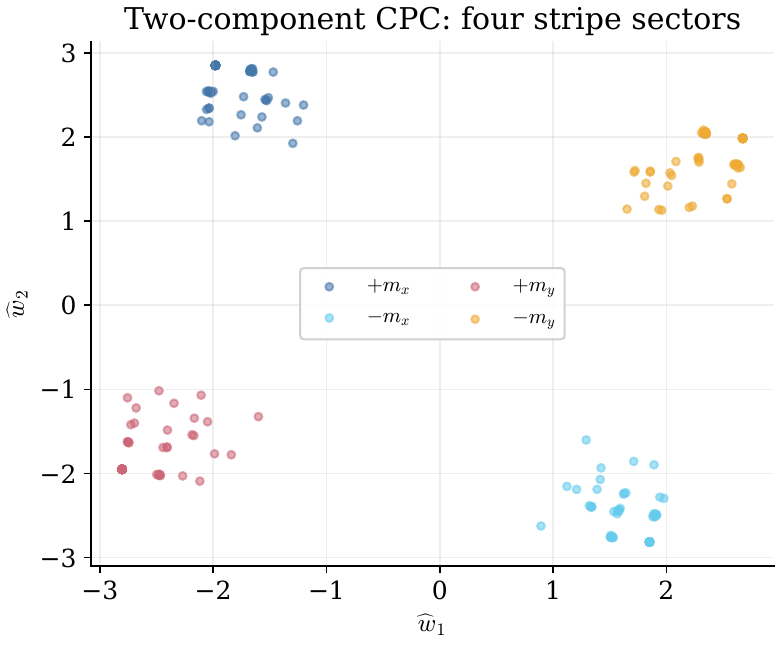}
 \caption{Collective predictive order ($K=2$) inferred from low-temperature
 stripe configurations at $T=1.2$, $1.4$, and $1.6$. Each point represents
 one test configuration. The separate four
 sectors associated with $+m_x$, $-m_x$, $+m_y$, and $-m_y$ are observed.
 Colors identify the conventional stripe sectors after learning;
 these labels were not supplied to CPC.}
 \label{fig:s3-stripe-sectors}
\end{figure}

Figure~\ref{fig:s3-stripe-sectors} shows the two components of the collective predictive order $\widehat w(S)$ inferred from test configurations at $T=1.2$, $1.4$, and $1.6$. The four distinguishable groups correspond to the positive and negative sectors of the two stripe orders, identified by color after learning. A polynomial map fitted to part of the test set reconstructs $(m_x,m_y)$ from the learned coordinate with $R^2=0.9861$ on the remaining configurations. 

At high temperature the masked prediction gain becomes negligible or slightly
negative: $\Delta\mathrm{NLL}\simeq-0.005$ for $K=2$ at $T=3.6$ means that
CPC predicts the hidden spins slightly worse than the independent-site
reference. A fixed two-component model is therefore
useful for the ordered ensemble, but does not prove that two predictive
components remain necessary in the disordered regime.

As a summary, the learned collective predictive order reproduces known existence of multiple order parameters, with their correct structure.


\subsection{Replica overlap in a spin glass}
\label{sec:s3-spin-glass}

A spin glass tests a different choice of local observations. We use the
fully connected Sherrington--Kirkpatrick (SK) model,
\begin{equation}
 H=-\sum_{i<j}J_{ij}s_i s_j,\qquad
 J_{ij}\sim\mathcal N(0,1/N),\qquad N=32.
 \label{eq:s3-sk}
\end{equation}
Every pair of spins interacts, irrespective of spatial separation. Thus
this infinite-range model has no specified spatial lattice dimension;
$N=32$ denotes the number of spins.
Two separate parallel-tempering ladders sample replicas with the same $J$
and temperature. Each agent now observes the four-state pair
$(s_i^{(1)},s_i^{(2)})$, rather than a single spin. The conventional
replica overlap is
\begin{equation}
 q_{12}=\frac1N\sum_i s_i^{(1)}s_i^{(2)}.
 \label{eq:s3-overlap}
\end{equation}
The choice to supply two replicas is a modeling input; their product and the
overlap formula are not supplied to the learning rule.

For scalar $w$ and the same Gaussian $p_0(w)$, replace the local probabilistic prediction model by
an unrestricted four-state distribution,
\begin{equation}
 p_{\theta^{(i)}}^{(i)}(u,v\mid w)
 =\frac{e^{b_{i,uv}+W_{i,uv}w}}
 {\sum_{u',v'=\pm1}e^{b_{i,u'v'}+W_{i,u'v'}w}},
 \qquad u,v=\pm1.
 \label{eq:s3-categorical}
\end{equation}
The regularization in Eq.~\eqref{eq:s3-regularization} now also sums over
the four states. Neither the parameters at different sites nor the four-state
weights are constrained to favor overlap. Data comprise ten replica pairs per
$(J,T)$ at eight temperatures from $0.55$ to $2.0$. The 44 independently
generated coupling realizations are split as 24/8/12 for
training/validation/test; hence every test $J$ is unseen during learning.
Sampling uses 700 initial sweeps and six sweeps between saved pairs.
Independent spin reversals and replica exchange are applied as exact
symmetries, with no site permutation. CPC receives neither $J$ nor temperature
labels. Integrals use 81 quadrature points on $[-4.2,4.2]$, and validation
selects $\lambda_W=1$ over $10$ with two initializations each.

To interpret the result, decompose the learned weights after training:
\begin{equation}
 W_{i,uv}=c_i+h_i^{(1)}u+h_i^{(2)}v+\kappa_iuv.
 \label{eq:s3-contrast}
\end{equation}
The common $c_i$ cancels from Eq.~\eqref{eq:s3-categorical}. Of the summed
squared norms of the other three contrasts, $98.55\%$ lies in $\kappa$.
Thus the overlap channel is selected by learning. 

\begin{figure}[t]
 \centering
 \includegraphics[width=0.60\textwidth]{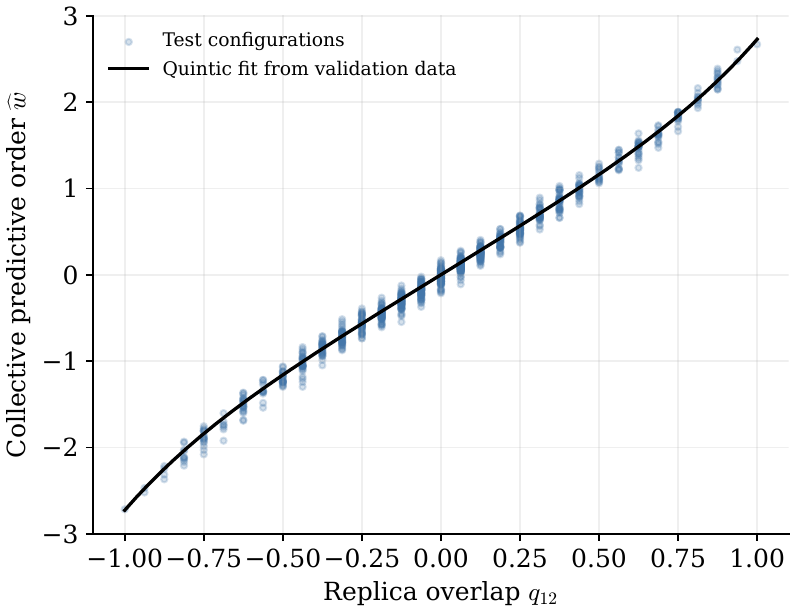}
 \caption{Relation between the replica overlap $q_{12}$ and the inferred
 collective predictive order $\widehat w$ in the fully connected SK model with $N=32$ spins. Each point represents one replica-pair configuration from the
 test set. The 960 configurations comprise ten samples at each of eight
 temperatures, $0.55\leq T\leq2.0$, for each of 12 independently generated
 realizations of the random couplings $\{J_{ij}\}$. The two replicas in
 each pair share the same couplings. The black curve is the odd-quintic map from $q_{12}$ to $\widehat w$, fitted using the separate validation set of coupling
 realizations and then held fixed. It gives $R^2=0.99200$ on the test
 configurations shown here.}
 \label{fig:s3-sk-map}
\end{figure}

Figure~\ref{fig:s3-sk-map} shows the relation between the learned
coordinate $\widehat w$ and the replica overlap $q_{12}$ for the test
configurations. To quantify this relation, we fit
$\widehat w \simeq a_1q_{12}+a_3q_{12}^3+a_5q_{12}^5$.
The coefficients are determined using configurations from the
validation set of coupling realizations, where each realization
is a separately generated set of random couplings $\{J_{ij}\}$.
Keeping these coefficients fixed, we evaluate the map on the test
set, whose coupling realizations were used neither for CPC training
nor for fitting the polynomial. The fit gives $R^2=0.99200$ on this
test set, showing that the learned coordinate is well described
as a nonlinear function of the replica overlap, even for previously
unseen coupling realizations.

For the second criterion, we perform the masked prediction test on the
12 test realizations of the random couplings, which were not used for
training or model selection.
Here the local observation is the pair of spins from the two replicas,
with four possible values. We reveal these pairs at $25\%$ of the sites
and predict the pairs at the remaining $75\%$. The independent-site
reference uses the training frequencies of the four local states.
At $T=0.55$, CPC reduces the prediction NLL relative to this reference by
$\Delta\mathrm{NLL}=0.1289$ per hidden site pair; the gain is also positive
for each of the 12 test coupling realizations evaluated separately.
Randomly shuffling the sites of the second replica in each test pair removes
this predictive improvement.

These calculations demonstrate the recovery of the replica
overlap across coupling realizations.
More generally, for all three examples the common result is concrete:
learning from configuration snapshots let the familiar macroscopic order parameter emerge, and their physical relevance is supported by prediction of observations withheld from inference. 
Although the examples depend on the chosen local probabilistic prediction model, observation structure, and candidate latent dimensions, our results are encouraging: the macroscopic order emerges through collective predictive coding.


\section{Emergent potential}
\label{sec:emergent-potential}

Section~\ref{sec:emergent-order} showed that the collective predictive order
$\widehat w(S)$ recovers familiar order parameters, up to a nonlinear change
of variable. We now ask whether its equilibrium fluctuations also reveal
the associated Landau potential. This extends the idea of
Sec.~\ref{sec:introduction}: after finding a macroscopic variable from
spin configurations, we seek the free energy that describes its preferred
values and their change with temperature. The free energy landscape should
respect hidden symmetry structure, if any, and in this sense, the symmetry can be 
emergent as the potential emerges.

We study the ferromagnetic and antiferromagnetic Ising models. The strategy 
for obtaining the emergent potential is summarized as follows. For each
model and lattice size, the parameters of Eq.~\eqref{eq:s3-local-model}
are learned once from training data configurations collected at several temperatures.
The resulting map $S\mapsto\widehat w(S)$ is then held fixed. Temperature
labels then enter at this stage to group the test configurations into thermal
ensembles. The frequencies of the inferred coordinates in each ensemble
then determine the potential. In the following, we define our emergent potential from CPC,
and present our results on the Ising models. Remarkably, Landau-like potentials
with phase transition are emergent, from which the Ising critical temperature 
is read accurately.

\subsection{Emergent Landau potential}
\label{sec:s4-constrained}

Here we shall define our potential only from the data configurations and the learned collective predictive order, as follows.

After learning the scalar map $\widehat w(S)$ of
Eq.~\eqref{eq:cpc-coordinate}, we apply it to equilibrium spin
configurations at each temperature. The inputs at this stage are the
learned map, the configurations, and their size and temperature labels.
For fixed $L,T$, let $S^{(r)}$, $r=1,\ldots,R$, denote these configurations.
Their collective predictive orders and measured distribution are
\begin{equation}
 P^{\mathrm{emp}}_{L,T}(\widehat w)
 \equiv \frac1R\sum_{r=1}^{R}\delta(\widehat w-\widehat w(S^{(r)})).
 \label{eq:s4-empirical}
\end{equation}
Thus each configuration contributes one value, and the fraction of values
in a given interval estimates the probability of that interval. We use
the test configurations of Sec.~\ref{sec:s3-prediction}, keeping all
learned parameters fixed. Grouping the configurations by temperature
allows us to follow the thermal evolution of this distribution.

Let $P_{L,T}(\widehat w)$ denote the underlying equilibrium distribution
estimated by Eq.~\eqref{eq:s4-empirical}. In terms of the equilibrium
probability $p_{L,T}(S)$ of a configuration, it is
\begin{equation}
 P_{L,T}(\widehat w)=\sum_S p_{L,T}(S)
 \delta\!\left(\widehat w-\widehat w(S)\right).
 \label{eq:s4-pushforward}
\end{equation}
The observed frequencies provide an estimate of this distribution directly.

We define the potential as
\begin{equation}
 \Phi_{L,T}(\widehat w) \equiv -k_{\rm B}T
 \log\frac{P_{L,T}(\widehat w)}{P_{L,T}^{\max}},\qquad
 P_{L,T}^{\max} \equiv \max_{\widehat w}P_{L,T}(\widehat w).
 \label{eq:s4-potential}
\end{equation}
This construction uses the relative frequencies of the inferred values.
For configurations sampled at thermal equilibrium, $\Phi_{L,T}$ is the
constrained free energy with its minimum set to zero.\footnote{
The thermodynamic identification follows by writing the equilibrium
probability as $p_{L,T}(S)=e^{-\beta H(S)}/Z_{L,T}$, with
$\beta=(k_{\rm B}T)^{-1}$ and $Z_{L,T}=\sum_S e^{-\beta H(S)}$.
The partition sum at fixed collective predictive order is
\begin{equation}
 \mathcal Z_{L,T}(\widehat w)
 =\sum_S e^{-\beta H(S)}\delta\!\left(\widehat w-\widehat w(S)\right),
 \qquad P_{L,T}(\widehat w)=\frac{\mathcal Z_{L,T}(\widehat w)}{Z_{L,T}}.
 \label{eq:s4-constrained-partition}
\end{equation}
Hence the free energy is $A_{L,T}(\widehat w)\equiv-k_{\rm B}T\log\mathcal Z_{L,T}(\widehat w)
=-k_{\rm B}T\log P_{L,T}(\widehat w)-k_{\rm B}T\log Z_{L,T}$. Equation~\eqref{eq:s4-potential}
is therefore $A_{L,T}(\widehat w)-\min_{\widehat w}A_{L,T}(\widehat w)$.
This identity supplies the thermodynamic interpretation; the numerical
construction uses the configuration frequencies in
Eq.~\eqref{eq:s4-empirical}.}
Frequently observed values form potential wells and are localized around the bottom of the potential, 
while rare intermediate values form
potential barriers. 

This gives the two stages of the construction a direct physical meaning.
Minimizing the dimensionless variational free energy $\mathcal F$ of
Eq.~\eqref{eq:cpc-total-F} selects the macroscopic variable $\widehat{w}(S)$.
Its equilibrium distribution then determines the constrained Landau
potential $\Phi_{L,T}(\widehat{w})$. 

Two notes on the potential follow.
First, the definition of the probability distribution Eq.~\eqref{eq:s4-empirical} is with a delta-function, thus it is quite spiky
and not suited for physical interpretation of the emergent potential. 
To obtain a smooth estimate from Eq.~\eqref{eq:s4-empirical}, we replace
each delta function by a normalized Gaussian of width $h_T$:
\begin{equation}
 \widehat P_{L,T}(\widehat w)=\frac{1}{R\sqrt{2\pi}h_T}
 \sum_{r=1}^{R}\exp\!\left[-\frac{(\widehat w-\widehat w(S^{(r)}))^2}{2h_T^2}\right].
 \label{eq:s4-kde}
\end{equation}
This is Gaussian kernel density estimation. 
We determine $h_T$ by leave-one-out likelihood cross-validation.\footnote{The leave-one-out likelihood cross-validation is a method to find an optimized value for the gaussian replacement. For each trial width, we remove one configuration and estimate the density from the remaining $R-1$ configurations. We evaluate this density
at the collective predictive order of the removed configuration and
repeat the procedure for every configuration. We choose the width that
maximizes the mean logarithm of these densities. This favors a width
for which the remaining configurations assign high probability density
to the values observed in the removed configurations.
We then impose the lower bound $h_T/w_{0,L}\geq0.012$ to suppress
fine structure associated with the numerical integration over $w$.}
We insert this estimate \eqref{eq:s4-kde}
into Eq.~\eqref{eq:s4-potential}. The figures show
\begin{equation}
 \beta\Phi_{L,T}(\widehat w)
 =-\log\frac{\widehat P_{L,T}(\widehat w)}
 {\max_{\widehat w}\widehat P_{L,T}(\widehat w)},
 \qquad \beta=(k_{\rm B}T)^{-1}.
 \label{eq:s4-estimated-potential}
\end{equation}
Thus the numerical procedure to obtain the emergent potential consists of evaluating the learned map on
each test configuration, smoothing the resulting distribution, and taking
its negative logarithm relative to the maximum density.

Second, note that the argument of our potential is not the magnetization but the collective predictive order of Sec.~\ref{sec:s3-local-model}, the latter is a nonlinear transform of the former. 
Let us look at the difference concerning the coordinates of the potential here.
For perfectly uniform or staggered weights, Eq.~\eqref{eq:s3-nonlinear-coordinate} gives
$\widehat w=G(A m)$, where $m$ is the corresponding magnetization and
$A$ is the proportionality factor between $h_W$ and $m$.
Writing this relation as $\widehat w=f(m)$, a smooth invertible $f$ gives
\begin{align}
 P_{\widehat w}(\widehat w)&=P_m(m(\widehat w))
 \left|\frac{dm}{d\widehat w}\right|,\nonumber\\
 \Phi_{\widehat w}(\widehat w)&=\Phi_m(m(\widehat w))+k_{\rm B}T
 \log\left|\frac{d\widehat w}{dm}\right|+C.
 \label{eq:s4-jacobian}
\end{align}
Here $C$ accounts for the separate choices of zero free energy.
The Jacobian therefore changes the potential shape and its polynomial
coefficients. Even the number of local minima can depend on a nonlinear
change of variable. In this paper, we examine the potential as a function of the collective predictive order.

\subsection{Emergent potential of Ising models and emergent symmetry}
\label{sec:s4-ising}

We apply this construction to the square-lattice Ising models of
Eq.~\eqref{eq:s3-ising-hamiltonians}, with periodic boundaries and
$|J|=k_{\rm B}=1$. Although the configurations and learned parameters
are separate from those of Sec.~\ref{sec:s3-ferro}, the newly learned weights align with the uniform pattern in the ferromagnetic model and the staggered pattern in the antiferromagnetic model. 
First here we describe the settings.
For each model and each $L=12,16,24,32,48,64$, a scalar
CPC model is learned independently from eleven temperatures between
$1.5$ and $3.5$. At each temperature, 70, 20 and 800 configurations are
allocated to training, validation and testing. Configurations are collected
from sixteen independently initialized Monte Carlo sequences using Wolff
cluster updates and are mixed before splitting into the three sets.
The spin orientation of each configuration is saved as produced by the
Monte Carlo updates.
Antiferromagnetic configurations are obtained from ferromagnetic ones by
$s_i\mapsto\epsilon_i s_i$, where $\epsilon_i=(-1)^{x_i+y_i}$. The roles of training, validation, and test data are the same as in
Sec.~\ref{sec:s3-prediction}.

\begin{figure}[tbp]
 \centering
 \includegraphics[width=\textwidth]{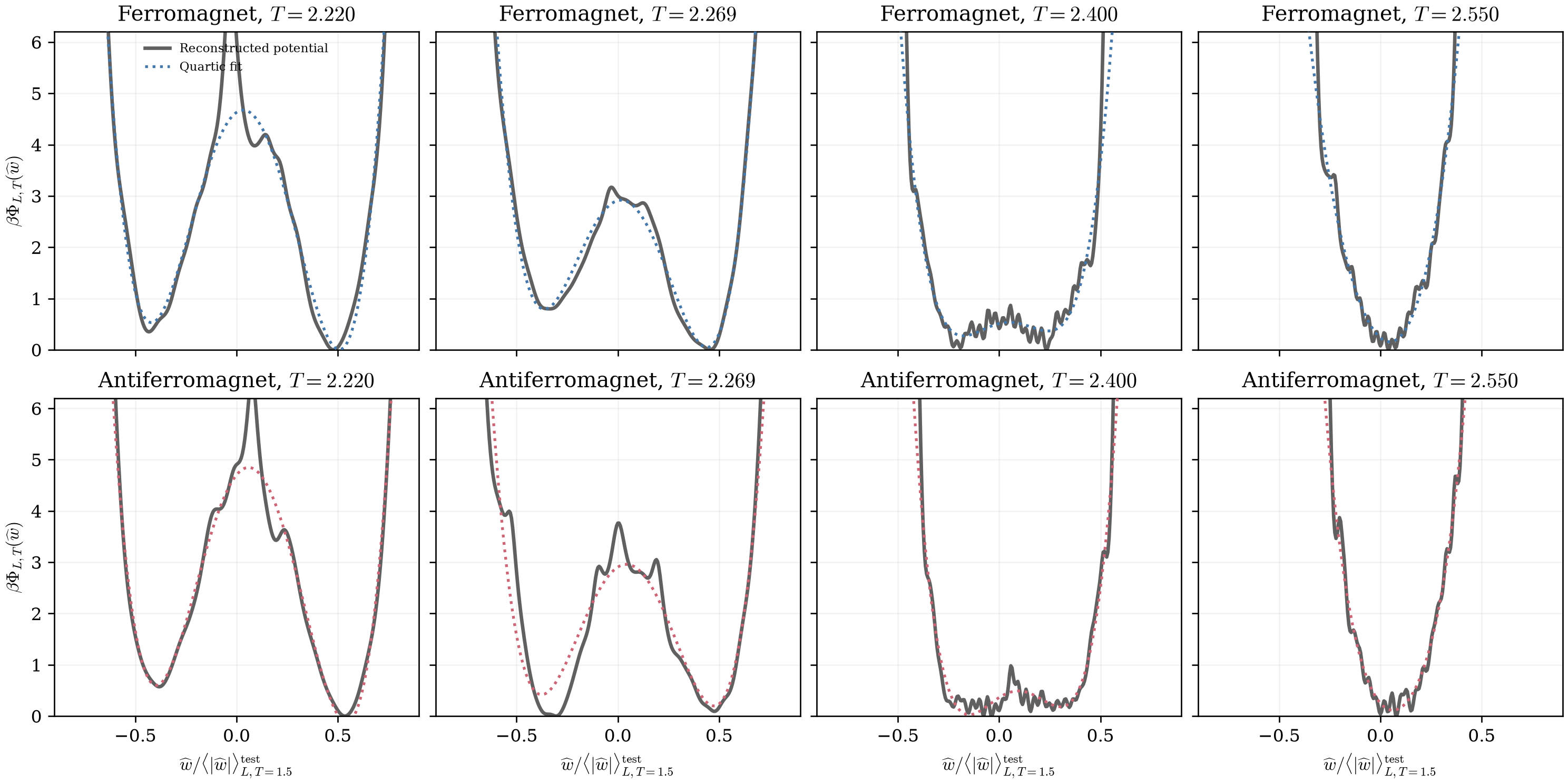}
 \caption{Emergent Landau potentials of the collective predictive order at
 $L=32$. Top: ferromagnet; bottom: antiferromagnet. From left to right,
 the temperatures are $T=2.22$, $T=T_c=2/\log(1+\sqrt2)\simeq2.269185$,
 $T=2.40$, and $T=2.55$, in units $|J|=k_{\rm B}=1$. Gray solid curves show
 the potentials reconstructed using Eq.~\eqref{eq:s4-kde}; colored dotted
 curves show the quartic fits of Eq.~\eqref{eq:s4-quartic}. The horizontal axis is
 $\widehat w/w_{0,L}$, where $w_{0,L}$ is defined in
 Eq.~\eqref{eq:s4-ising-scale}, and each reconstructed
 potential has its minimum set to zero. Fit coefficients are listed in
 Table~\ref{tab:s4-quartic-coefficients}.}
 \label{fig:s4-quartic-profiles}
\end{figure}

Our results of the emergent potentials are presented in Fig.~\ref{fig:s4-quartic-profiles}.  
We use $L=32$ to illustrate the reconstruction.
The gray solid curves in
Fig.~\ref{fig:s4-quartic-profiles} show the potentials reconstructed directly from the measured coordinates. 
This captures the well-known behavior of the Landau potential for magnetization in the Ising models.

It is quite intriguing that in fact the potentials approximately preserve $Z_2$ symmetry $\widehat w \leftrightarrow -\widehat w$. We like to remind the readers that we have not told to the model the symmetry of the Hamiltonian. The CPC method automatically reconstruct the Landau potential through the automatically discovered order parameter, and the resultant potentials retain the symmetry.

Furthermore, the tendency of the spontaneous symmetry breaking can be apparently seen in the temperature dependence of the emergent potentials. With the high temperature samples, the reconstructed potential is just a single well, while at low temperature, it has two wells. There is a phase transition between them.
The two wells near the transition approach the origin on heating and give way to a single well near the
origin at $T=2.55$. 

Here are some detailed notations and numbers used for obtaining the potentials in Fig.~\ref{fig:s4-quartic-profiles}.
For comparisons between models and sizes, we express the horizontal axis
in units of the mean low-temperature amplitude,
\begin{equation}
 w_{0,L}\equiv\langle|\widehat w(S)|\rangle_{L,T=1.5}^{\rm test}
 =\frac1R\sum_{r=1}^{R}|\widehat w(S^{(r)})|.
 \label{eq:s4-ising-scale}
\end{equation}
The sum uses the $R=800$ test configurations at $T=1.5$ for each model
and size. Each plot uses $\widehat w/w_{0,L}$ on the horizontal axis,
with $w_{0,L}$ held fixed throughout the temperature scan.
A constant rescaling multiplies the probability density by a constant,
which cancels in the ratio defining $\Phi_{L,T}$ in
Eq.~\eqref{eq:s4-potential}.

\subsection{Landau coefficients and critical temperature estimate}
\label{sec:s4-quartic}

The reconstructed emergent Landau potential shown in Fig.~\ref{fig:s4-quartic-profiles} actually exhibits 
a phase transition, then can we pin down the critical temperature from the potentials?
The exact thermodynamic critical temperature is $T_c=2/\log(1+\sqrt2)=2.269185\ldots$, while Fig.~\ref{fig:s4-quartic-profiles} suggests that the transition temperature is higher than that. 
In the following, we study the finite-size effect on this model and consider the infinite volume limit to numerically pin-down the value of the critical temperature.

At the reference size $L=32$, we describe the change of shape in familiar
Landau terms by fitting the measured potentials at each temperature to
\begin{equation}
 \beta\Phi_{L,T}(\widehat w)\simeq
 \sum_{n=0}^{4}a_n(L,T)
 \left(\frac{\widehat w}{w_{0,L}}\right)^n.
 \label{eq:s4-quartic}
\end{equation}
The coefficients $a_n$ refer to powers of the ratio
$\widehat w/w_{0,L}$ used in the figures. Equivalently, the coefficient
of $\widehat w^n$ is $a_n/w_{0,L}^n$. Here $\Phi$ has its minimum set to
zero. The least-squares fit uses
points with $|\widehat w|/w_{0,L}\leq0.9$ and $\beta\Phi\leq6$, avoiding regions where very few configurations are observed. We retain the odd terms to describe the asymmetry of the
measured profiles. We report coefficients for $T\leq2.55$: at higher temperatures the observed
range of $\widehat w$ becomes too narrow to constrain the quartic term reliably.
Figure~\ref{fig:s4-quartic-profiles} also shows representative fits at $L=32$;
their coefficients are listed in Table~\ref{tab:s4-quartic-coefficients}.

\begin{table}[tbp]
 \centering
 \caption{Representative quartic coefficients at $L=32$, in the coordinate
 and potential convention of Eqs.~\eqref{eq:s4-ising-scale} and
 \eqref{eq:s4-quartic}. F and AF denote the ferromagnet and antiferromagnet.
 Digits are rounded to reflect the sampling uncertainty.}
 \label{tab:s4-quartic-coefficients}
 \begin{tabular}{ccrrrrr}
 \hline
 Model & $T$ & $a_0$ & $a_1$ & $a_2$ & $a_3$ & $a_4$\\
 \hline
 F  &2.220&4.6& 2.7&$-40$&$-15$& 95\\
 F  &2.269&2.9& 1.0&$-30$&$-12$& 96\\
 F  &2.400&0.5& 1.0& $-9$&$-17$&112\\
 F  &2.550&0.2&$-3.3$&35&10&59\\
 AF &2.220&4.7&4.8&$-39$&$-26$&95\\
 AF &2.269&2.9&2.4&$-28$&$-15$&81\\
 AF &2.400&0.4&2.5&$-9$&$-49$&149\\
 AF &2.550&0.2&$-4.5$&40&$-45$&137\\
 \hline
 \end{tabular}
\end{table}

The numerical values in Table~\ref{tab:s4-quartic-coefficients} show that $a_2$ changes from
negative to positive while $a_4$ remains positive at the listed temperatures. This signifies the phase transition described in the emergent Landau potential.
To quantify this sign change numerically, we fit the quadratic coefficient over
$2.22\leq T\leq2.55$ to a straight line,
\begin{equation}
 a_2(L,T)\simeq A_L[T-T_*(L)].
 \label{eq:s4-a2-crossing}
\end{equation}
Here $A_L$ is the fitted slope and $T_*(L)$ is the temperature at which
that line reaches zero. We use this definition at every size, and numerically evaluate them to build Table \ref{tab:s4-crossover-sizes}.


\begin{table}[tbp]
 \centering
 \caption{Temperatures $T_*(L)$ at which the fitted
 quadratic Landau coefficient vanishes. Brackets give the 95\% intervals from
 300 resamplings of the test configurations.}
 \label{tab:s4-crossover-sizes}
 \begin{tabular}{ccccc}
 \hline
 &\multicolumn{2}{c}{Ferromagnet}&\multicolumn{2}{c}{Antiferromagnet}\\
 $L$&$T_*(L)$&95\% interval&$T_*(L)$&95\% interval\\
 \hline
12 & 2.512 & [2.487, 2.595] & 2.493 & [2.467, 2.565]\\
16 & 2.513 & [2.479, 2.592] & 2.518 & [2.485, 2.633]\\
24 & 2.485 & [2.458, 2.560] & 2.484 & [2.451, 2.521]\\
32 & 2.413 & [2.393, 2.435] & 2.406 & [2.386, 2.431]\\
48 & 2.356 & [2.345, 2.372] & 2.345 & [2.338, 2.358]\\
64 & 2.336 & [2.304, 2.355] & 2.337 & [2.328, 2.353]\\
 \hline
 \end{tabular}
\end{table}

\begin{figure}[tbp]
 \centering
 \includegraphics[width=\textwidth]{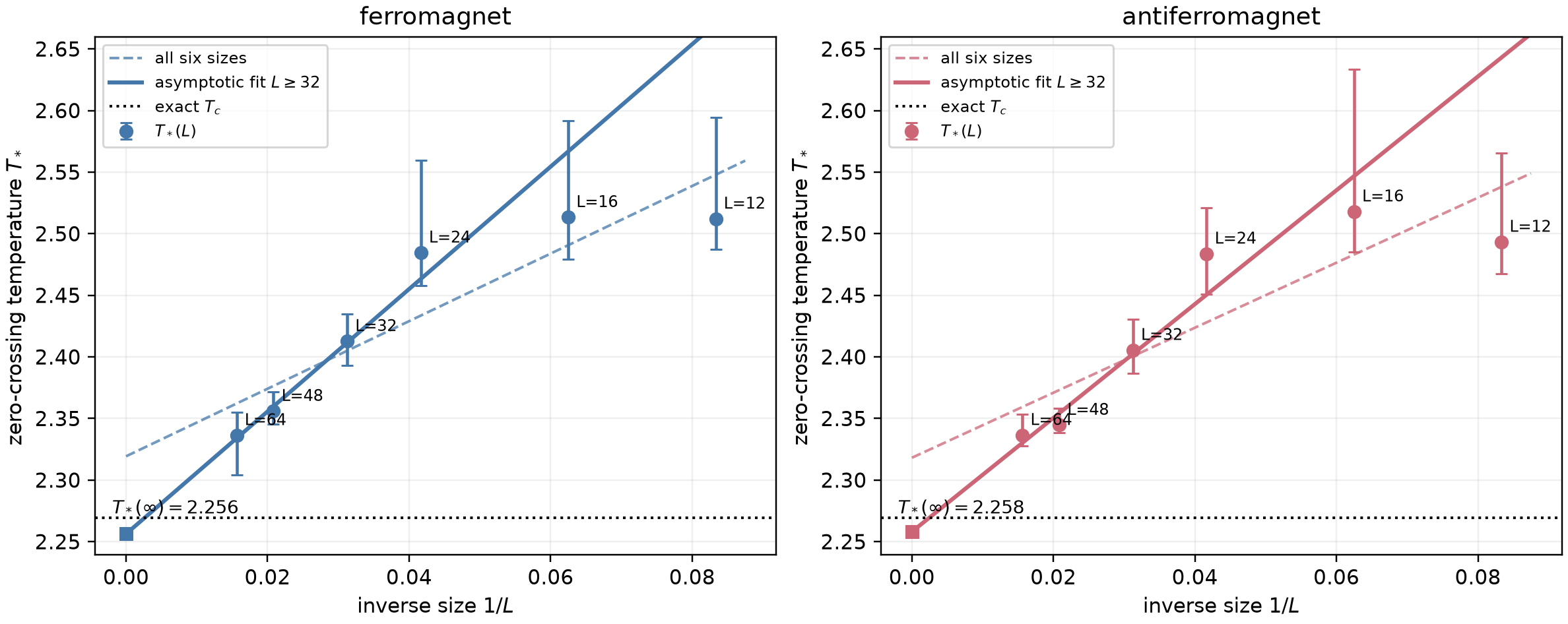}
 \caption{Finite-size extrapolation of $T_*(L)$, the temperature at which
 the linear fit to $a_2(T)$ vanishes. Points show the six system sizes;
 error bars are 95\% intervals from resampling test configurations. Dashed lines fit
 all six sizes, while solid lines fit only $L=32,48,64$ using
 Eq.~\eqref{eq:s4-tc-extrapolation}. Squares at $1/L=0$ mark the latter
 extrapolations. The horizontal dotted line is the exact
 $T_c=2.269185\ldots$.}
 \label{fig:s4-tc-extrapolation}
\end{figure}

A standard finite-size estimate of the transition shift is obtained by
requiring the correlation length $\xi\propto|T-T_c|^{-\nu}$ to become
comparable to $L$. This suggests a shift proportional to $L^{-1/\nu}$.
We test the simplest leading form
\begin{equation}
 T_*(L)=T_*(\infty)+\frac{c}{L},
 \label{eq:s4-tc-extrapolation}
\end{equation}
where $c$ and $T_*(\infty)$ are fitted constants.\footnote{
For the two-dimensional Ising model, the known value is $\nu=1$, thus this fitting function is rather natural. 
Instead we could fit $\nu$ numerically from the data too.}
Figure~\ref{fig:s4-tc-extrapolation} compares least-squares fits with
equal weight for each size, using either all six sizes or only
$L=32,48,64$. The latter reduces the influence of smaller systems and thus is adopted here.
Its intercept at $1/L=0$ gives
\begin{equation}
 T_*^{\mathrm F}(\infty)=2.256,\qquad
 T_*^{\mathrm{AF}}(\infty)=2.258.
 \label{eq:s4-tc-estimates}
\end{equation}
Repeating the extrapolation for the resampled data gives 95\% intervals
$[2.203,2.300]$ and $[2.223,2.298]$, respectively. Both contain the exact value 
$T_c=2.269185\ldots$.

The physical result is that the collective predictive order found from
spin observations supports a Landau description: its measured equilibrium
potential develops two wells, which approach a single central well on
heating. The temperature extracted from the quadratic coefficient has a
finite-size shift whose extrapolation is consistent with the known
transition. Together with Sec.~\ref{sec:emergent-order}, this establishes
the two steps of the construction: CPC selects the macroscopic variable from the mixed-temperature configurations,
and it also derives its Landau potentials from fixed-temperature statistics.


\section{Agentic communication between atoms}
\label{sec:agentic-communication}

Sections~\ref{sec:emergent-order} and \ref{sec:emergent-potential} identified macroscopic order and its potential from spin configurations. Their inference step combined all site observations in a numerical integral, that is, the information of the observation at all the sites is collected and computed for the inference, which is analogous to the existence of a central computer. On the other hand, for example the emergence of language or symbols in human history does not need such a central computer for the inference. Rather, communication between agents, not the centralized communication, can produce the language and symbols as a common variable. In fact, the virtue of CPC is that
 this kind of decentralized communication can also actually attain sharing of a common emergent variable \cite{Taniguchi:2023MHNG,Inukai:2023RMHNG}. 

Thus here in this section we ask whether a macroscopic order parameter can be learned through communication between individual sites. Each site proposes or evaluates a common variable using its own observed spin, and learns its own parameters from the outcome. This follows the CPC picture of a shared name acquiring meaning through communication, and establishes the Atom-as-an-Agent principle.

The relevant procedure constructed in the scheme of CPC is called Metropolis--Hastings naming game (MHNG), and it  provides the basic rule \cite{Taniguchi:2023MHNG,Inukai:2023RMHNG}: one agent proposes a name,
and another accepts it according to how well it explains that agent's
observation. Here the name is a real number $w$. The acceptance step is
familiar from Monte Carlo physics, with a ratio of local observation
probabilities playing the role of a Boltzmann-weight ratio.
Repeating this communication and acceptance/rejection process many times for the whole community of agents, all the agents can successfully obtain a shared notion. For the case of two agents, the MHNG communication rule has been shown to have the same stationary distribution as centralized inference of the shared variable, with the observations and model parameters held fixed \cite{Taniguchi:2023MHNG,Inukai:2023RMHNG}.

\begin{figure}[tbp]
 \centering
 \includegraphics[width=\textwidth]{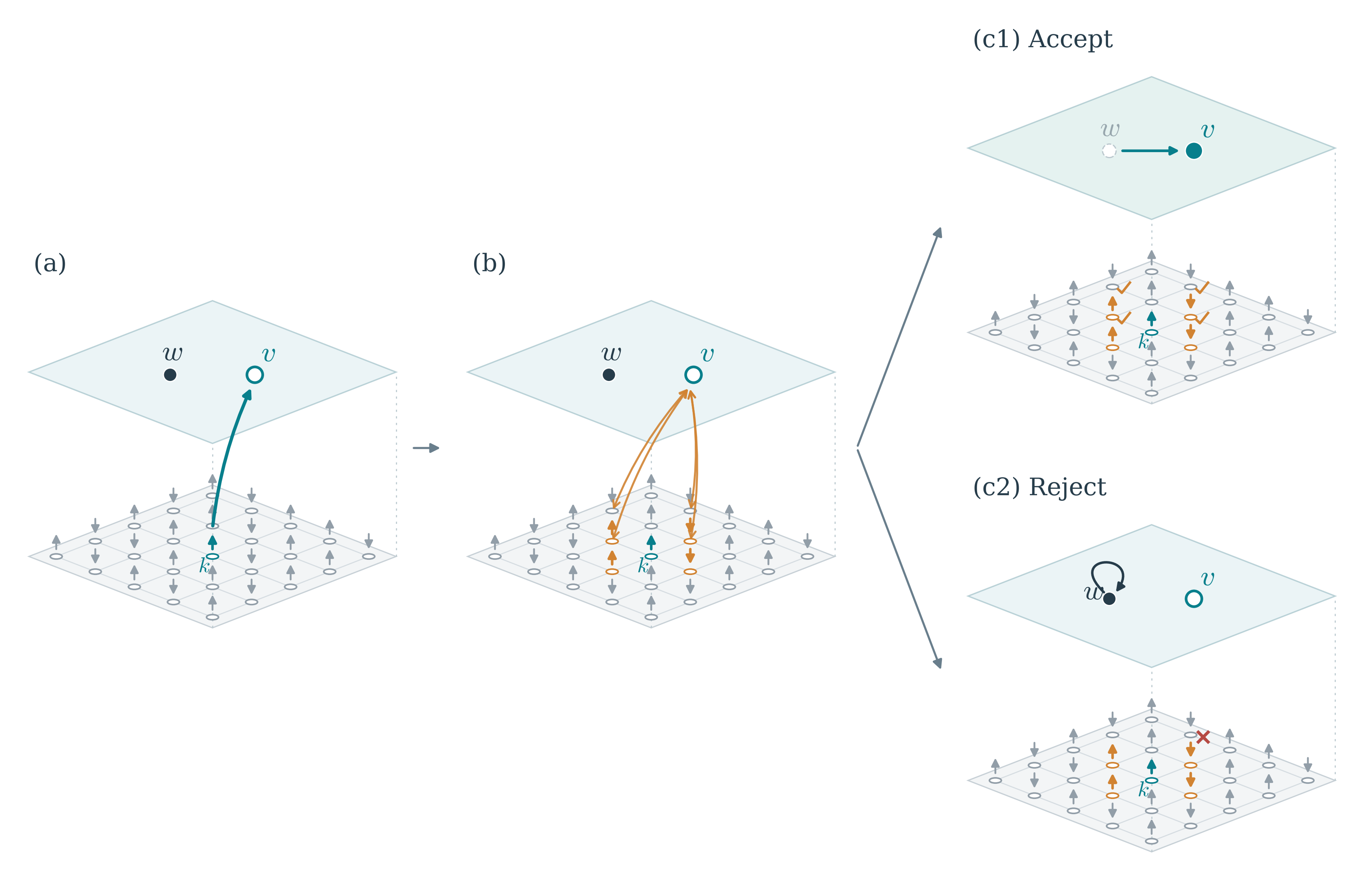}
 \caption{One communication step of the four-neighbor MHNG.
 The lower layer shows a fixed observed spin configuration; the upper
 layer represents values of the single variable shared by all sites.
 (a) Site $k$ (teal) proposes $v$ from its local conditional distribution,
 Eq.~\eqref{eq:s5-speaker}. (b) Its four nearest neighbors (orange)
 independently compare $v$ with the current value $w$ using
 Eq.~\eqref{eq:s5-listener}; orange links depict these local evaluations.
 (c1) All four listeners accept (check marks), and the shared state changes
 from $w$ to $v$. (c2) At least one listener rejects (cross), and the
 shared state remains $w$, as indicated by the loop arrow.
 Panels (c1) and (c2) are alternative outcomes. Filled circles in the
 upper layer indicate the current shared value. Both outcomes count as
 one communication step, after which a new proposing site is chosen.
 The spin configuration stays fixed throughout.}
 \label{fig:s5-mhng-schematic}
\end{figure}

In this section, we demonstrate that learning through four-neighbor MHNG recovers the same ferromagnetic ordering pattern as centralized CPC in Sec.~\ref{sec:s3-ferro}.

\subsection{A shared variable proposed and tested by individual spins}
\label{sec:s5-protocol}

We first specify how individual sites propose and assess a common
explanation of their spin configuration. We use the scalar model of
Eq.~\eqref{eq:s3-local-model}. For configuration $S^{(a)}$, define the
probability assigned by site $i$ to its observed spin:
\begin{equation}
 L_i^{(a)}(w)\equiv
 \frac{\exp[s_i^{(a)}(b_i+W_iw)]}{2\cosh(b_i+W_iw)},
 \qquad p_0(w)=\frac{e^{-w^2/2}}{\sqrt{2\pi}}.
 \label{eq:s5-local-model}
\end{equation}
The two parameters $(W_i,b_i)$ are learned separately at each site.
A positive $W_i$ associates a positive common $w$ with an increased
probability of an up spin; a negative $W_i$ gives the opposite association.
Learning these associations gives the common variable its meaning.

Hold one observed configuration and the parameters fixed, and suppress
$a$ for the following communication step. The lower layer in
Fig.~\ref{fig:s5-mhng-schematic} shows the observed spins; the upper
layer represents the single shared variable. Starting from its current
value $w$, choose a proposing site $k$ uniformly from all $N$ sites.
\begin{description}
\item[\underline{Step (a)} \hspace*{5mm}]
In Fig.~\ref{fig:s5-mhng-schematic} (a), this site $k$ uses its own observation to draw a candidate $v$ from the following normalized distribution at $k$,
\begin{equation}
 g_k(v)\equiv p_0(v)L_k(v)\Big{/}\int dv\,p_0(v)L_k(v).
 \label{eq:s5-speaker}
\end{equation}
By Bayes' theorem, this is the distribution of the common variable
conditioned on that site's spin alone.\footnote{It can be sampled using only local
probabilities: draw $v$ from the unit Gaussian and a uniform random number
$z\in[0,1]$; keep $v$ if $z<L_k(v)$, and repeat the draws otherwise.
The retained candidates have density $g_k$.}
\item[\underline{Step (b)}\hspace*{5mm}]
In Fig.~\ref{fig:s5-mhng-schematic}(b), each of the four nearest
neighbors $j\in\mathcal N(k)$ independently accepts the candidate with
probability
\begin{equation}
 A_j(w,v)\equiv \min\!\left(1,\frac{L_j(v)}{L_j(w)}\right).
 \label{eq:s5-listener}
\end{equation}
A candidate that better explains the listener's spin is always accepted;
a less probable explanation is accepted with the indicated probability, that is the Metropolis-Hastings test.
All four listeners compare the same current value $w$ with the same candidate $v$. 
\item[\underline{Step (c)}\hspace*{5mm}]
If all agree, the shared state becomes $v$, as shown in
panel (c1). If any listener rejects, the state remains $w$, represented by
the loop in panel (c2). These panels show alternative outcomes of the
same decision. The probability of acceptance, conditional on the speaker, is
\begin{equation}
 A_k^{(4)}(w,v)\equiv \prod_{j\in\mathcal N(k)}
 \min\!\left(1,\frac{L_j(v)}{L_j(w)}\right).
 \label{eq:s5-acceptance}
\end{equation}
\end{description}
Every proposal counts as one communication step, including a rejection,
which contributes a repeated current value to the sequence. Local
proposals and decisions thus generate successive shared explanations
of the configuration, while the observed spins remain fixed.

\subsection{Numerical steps for learning}
\label{sec:s5-learning}

We now construct a learning procedure in which each site adjusts its
association with the shared variable, starting from random weights.
We use the saved ferromagnetic Ising configurations of
Sec.~\ref{sec:s3-ferro}: $L=32$, $N=1024$, and thirteen temperatures
from $1.5$ to $3.5$, including $T_c$. At each temperature there are
240 training, 60 validation, and 200 test configurations, giving
$M=3120$, 780, and 2600 configurations in total.\footnote{The saved configurations were generated by Wolff updates, with 40 initial sweeps and one sweep between saved configurations; a random
global spin reversal was applied on saving. A sweep accumulates
cluster flips until the number of flipped spins reaches $N$. The three
data sets use different saved indices from the same Monte Carlo sequences.
Temporal correlations of these configurations would require a separate
statistical assessment.}
The training set combines all temperatures. Learning uses each site's spin observations and the
periodic square lattice's nearest-neighbor relations for communication.

Each run starts from independently generated Gaussian numbers $\xi_i$,
with
\begin{equation}
 W_i^{(0)}=0.08\frac{\xi_i}{\operatorname{std}_j\xi_j},\qquad
 b_i^{(0)}=\operatorname{atanh}\!\left[
 \operatorname{clip}\!\left(\frac1M\sum_a s_i^{(a)},-0.95,0.95\right)
 \right].
 \label{eq:s5-initialization}
\end{equation}
Here $\operatorname{std}$ is the sample standard deviation, and
$\operatorname{clip}(x,l,u)$ restricts $x$ to the interval $[l,u]$.
Thus the initial weights have random signs, and each initial bias is
set by that site's observed mean spin. Each configuration starts with
an independent $w^{(a)}$ drawn from the unit Gaussian.

One learning epoch consists of communication followed by parameter
updates. For each fixed spin configuration, perform 64 communication steps,
then 256 further steps, recording the current shared value $w$
after every 32 steps. 
This gives $D=8$ samples $w_{ad}$ for each configuration $S^{(a)}$,
where $d=1,\ldots,D$ labels the recorded values.
Each site uses these samples to average over the fluctuations of
the shared variable and updates its own parameters to reduce
\begin{equation}
 Q(\theta) \equiv \frac1D\sum_{a,d,i}
 \left[\log 2\cosh(b_i+W_iw_{ad})
       -s_i^{(a)}(b_i+W_iw_{ad})\right]
 +\Omega_{\mathrm{reg}}(\theta),
 \label{eq:s5-Q}
\end{equation}
with the regularization of Eq.~\eqref{eq:s3-regularization}.\footnote{
This sum separates into independent two-parameter problems, one at each
site. For example, the direction of decreasing $Q$ in $W_i$ is
\begin{equation}
 -\frac{\partial Q}{\partial W_i}
 =\frac1D\sum_{a,d}
 \left[s_i^{(a)}-\tanh(b_i+W_iw_{ad})\right]w_{ad}
 -\lambda_W W_i.
 \label{eq:s5-local-gradient}
\end{equation}
It compares the observed spin with the site's prediction for the shared
value. The corresponding update of $b_i$ uses the same difference,
with the factor $w_{ad}$ replaced by one and regularization
$-\lambda_b b_i$.}
The parameter update proceeds as follows. With the communication
samples fixed, each site takes four Newton steps to reduce $Q$ with
respect to its own parameters $(W_i,b_i)$.\footnote{The nonlinear sample averages used in these updates are evaluated
using interpolation with grid spacing $0.01$.}
To avoid excessively large changes, each increment is restricted
to $[-0.25,0.25]$, and the parameters after the four steps are
averaged with their values before the update.

After updating the parameters, communication for each configuration
$S^{(a)}$ resumes from its last recorded value $w_{aD}$ in the next epoch.

For model selection, we partly use the centralized computation, as follows. 
We use the same regularization candidates $\lambda_W=1,10,100$ and validation-likelihood criterion as in Sec.~\ref{sec:s3-ferro}. Here all three initializations for each $\lambda_W$ are random, and each run lasts 60 epochs. For each $\lambda_W$, we retain the run with the smallest final training objective $\mathcal J$; validation likelihood then selects $\lambda_W=1$. Note that these selection scores are evaluated by centralized
one-dimensional integration over $w$, thus we use centralized computation to choose among the trained models, while the samples used for their parameter updates are generated by MHNG communication.

We next examine the collective predictive order obtained by the MHNG algorithm.

\subsection{The order parameter obtained through communication}
\label{sec:s5-order}

We test whether the learned common variable represents ferromagnetic
order by examining the spatial weights and comparing its inferred value
with magnetization on test configurations.

\begin{figure}[tbp]
 \centering
 \includegraphics[width=0.85\textwidth]{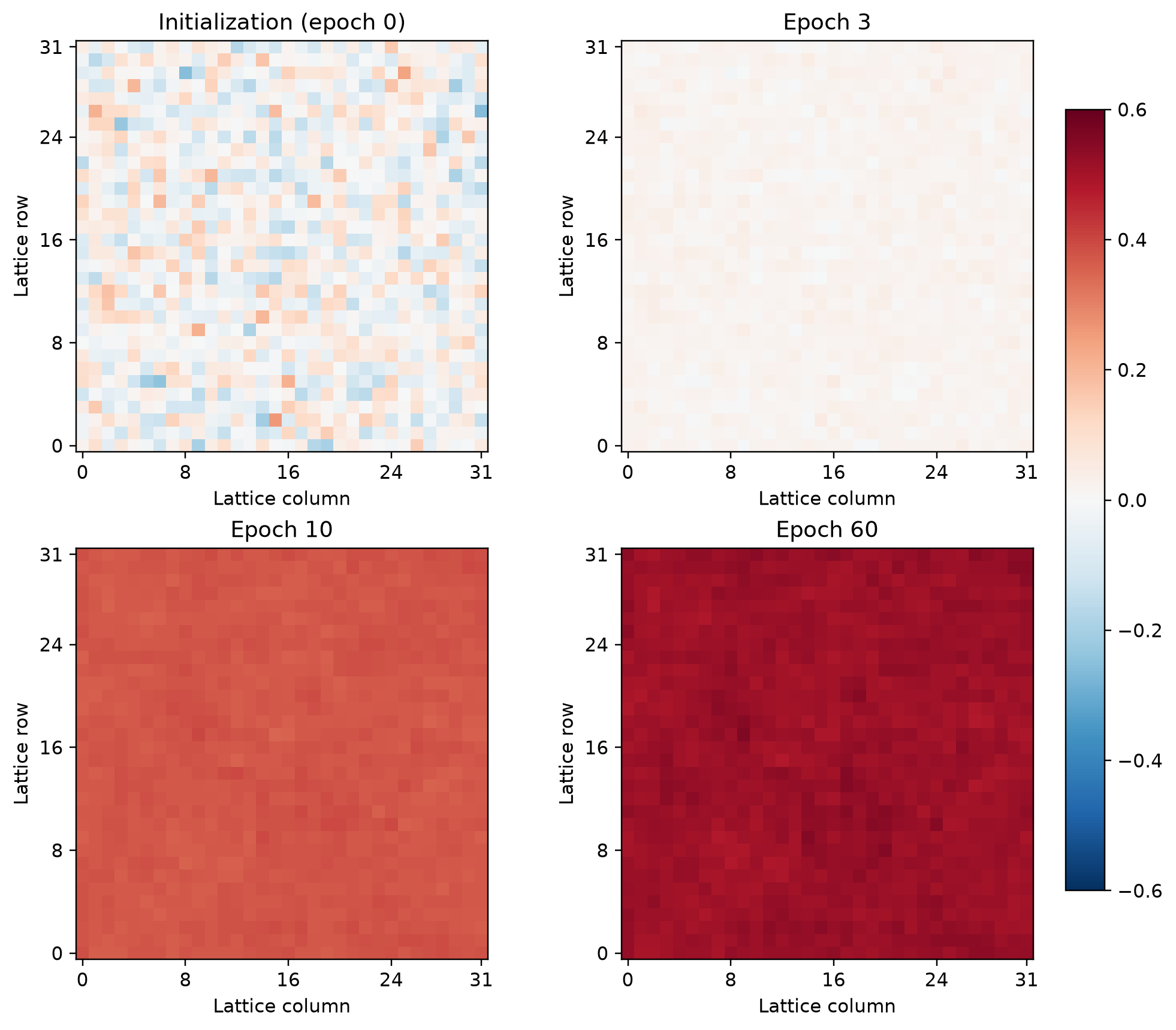}
 \caption{Site weights $W_i$ learned with four-neighbor communication,
 at initialization and epochs 3, 10, and 60. All panels use the same
 color scale $[-0.6,0.6]$ and the same overall sign convention.
 Random initial signs evolve into a nearly uniform pattern.}
 \label{fig:s5-nearest4-weights}
\end{figure}

See Figure~\ref{fig:s5-nearest4-weights} which shows the evolution of the selected
model's weights. Their normalized overlap with the uniform pattern,
$|\sum_iW_i|/\sqrt{N\sum_iW_i^2}$, increases from $0.01033$ to $0.999726$.
The other eight runs reach overlaps between $0.999708$ and $0.999748$.
After choosing one common overall sign for display, the final weights
have mean $0.515033$ and spatial standard deviation $0.012052$.
Thus learning from local decisions recovers a nearly uniform coupling
of the shared variable to the spins.

To evaluate the order encoded in the learned local probabilistic prediction models, we hold
$(W_i,b_i)$ fixed and infer the shared variable separately at each site.
For a new spin configuration $S$, site $i$ assigns the proposal
distribution of Eq.~\eqref{eq:s5-speaker} to $v$. We compute its mean
and then average over sites:
\begin{equation}
 \mu_i(s_i)\equiv
 \frac{\int dv\,v\,p_0(v)L_i(v)}{\int dv\,p_0(v)L_i(v)},
 \qquad
 \bar w_{\mathrm{loc}}(S)\equiv\frac1N\sum_{i=1}^{N}\mu_i(s_i).
 \label{eq:s5-readout}
\end{equation}
Each local mean uses only that site's observed spin and learned
parameters. The two values $\mu_i(s_i=\pm1)$ are computed once per site
using 80-point Gauss--Hermite quadrature. 
We use $\bar w_{\mathrm{loc}}$ as a local readout
of collective predictive order, distinguished from the full conditional
mean $\widehat w$ of Eq.~\eqref{eq:cpc-coordinate}.

\begin{figure}[tbp]
 \centering
 \includegraphics[width=\textwidth]{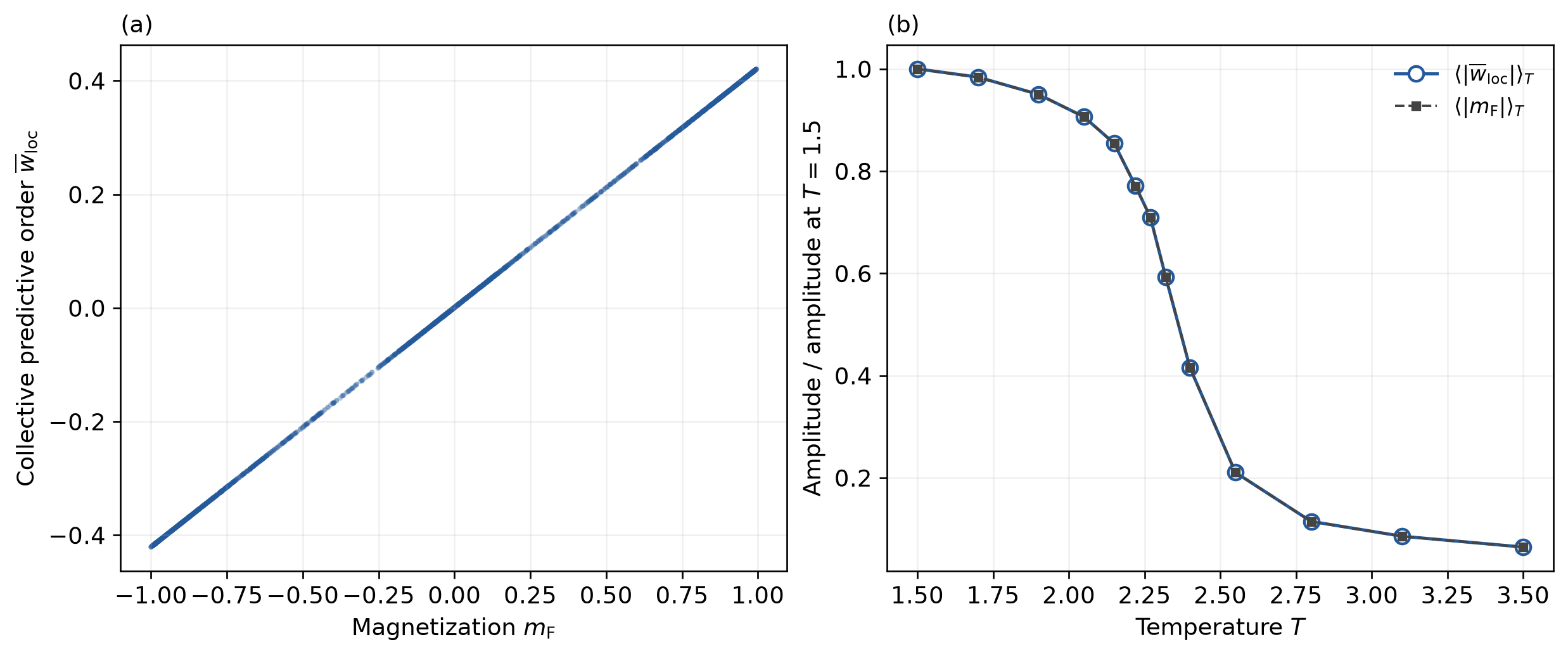}
 \caption{Collective predictive order evaluated by local inference after
 learning through four-neighbor MHNG. (a) The spatial average
 $\bar w_{\mathrm{loc}}(S)$ of the local proposal means versus
 magnetization on 2600 test configurations. A common sign is chosen
 so that the mean learned weight is positive.
 (b) Temperature dependence of $\langle|\bar w_{\mathrm{loc}}|\rangle_T$
 (open circles, solid line) and $\langle|m_{\mathrm F}|\rangle_T$
 (squares, dashed line), each normalized by its own value at $T=1.5$.
 Each temperature has 200 test configurations; the two curves almost
 overlap.}
 \label{fig:s5-nearest4-order}
\end{figure}

Figure~\ref{fig:s5-nearest4-order} shows that the spatial average $\bar w_{\mathrm{loc}}$ of
the local inference means closely follows magnetization, with
$|\operatorname{corr}(\bar w_{\mathrm{loc}},m_{\mathrm F})|=0.9999996$
on the 2600 test configurations. Thus the interpretation of this collective predictive order is the magnetization.

This nearly linear relation, as opposed to the results in the previous sections, reflects
both the learned weights and the definition of the readout.
Since each spin takes only two values, each local mean can be written as
\begin{equation}
 \mu_i(s_i)=
 \frac{\mu_i(+1)+\mu_i(-1)}{2}
 +\frac{\mu_i(+1)-\mu_i(-1)}{2}\,s_i.
\end{equation}
Thus $\bar w_{\mathrm{loc}}$ is a weighted sum of spins plus a constant.
For uniform weights and zero biases, symmetry gives
$\mu_i(s_i)=c\,s_i$ with a common constant $c$, so that
$\bar w_{\mathrm{loc}}=c\,m_{\mathrm F}$.
The learned model is close to this limit.

This readout tests whether the local probabilistic prediction models learned through MHNG
have acquired the common ferromagnetic pattern. Its linear form
differs from the nonlinear relation obtained by combining all
sites' probabilities before taking the conditional mean
$\widehat w$ in Sec.~\ref{sec:s3-ferro}.
The mean absolute values of $\bar w_{\mathrm{loc}}$ and $m_{\mathrm F}$
also have nearly identical temperature dependence after normalization
by their respective values at $T=1.5$, as shown in Figure~\ref{fig:s5-nearest4-order}.
Table~\ref{tab:s5-readout} lists the unnormalized amplitudes.

\begin{table}[tbp]
 \centering
 \caption{Mean absolute local readout and magnetization on test
 configurations, with 200 configurations at each temperature.}
 \label{tab:s5-readout}
 \begin{tabular}{crr}
 \hline
 $T$ & $\langle|\bar w_{\mathrm{loc}}|\rangle_T$
     & $\langle|m_{\mathrm F}|\rangle_T$\\
 \hline
 $1.5$ & 0.41602 & 0.98661\\
 $T_c$ & 0.29536 & 0.70033\\
 $3.5$ & 0.02678 & 0.06350\\
 \hline
 \end{tabular}
\end{table}

These results give the Atom-as-an-Agent principle a concrete realization
in a system of spins. Each site proposes or evaluates a shared explanation
using its own observation, and learns how that explanation relates to its
local state. Starting from random weights, these exchanges produce a
common variable that represents ferromagnetic order. The variable is
shared across the system, while proposal, acceptance, and parameter
updates are assigned to individual sites. In this construction,
macroscopic order acquires a predictive meaning through the communication
and learning of its microscopic constituents.\footnote{In this section we study the recovery of
the same physical ordering pattern, but we do not require the four-neighbor communication rule to reproduce the probability distribution of centralized inference. The MHNG for two agents is shown to reproduce the same distribution in \cite{Taniguchi:2023MHNG,Inukai:2023RMHNG}, while our case generalizes the situation to four-neighboring agents. More concretely,  for fixed spin configuration and model parameters, our rule generates
a stationary distribution $q_4(w\mid S)$, which generally differs from
$\pi(w\mid S)\propto p_0(w)\prod_i L_i(w)$.
For example, if all local likelihoods equal $L(w)$, then
$q_4(w\mid S)\propto p_0(w)L(w)^5$, whereas
$\pi(w\mid S)\propto p_0(w)L(w)^N$.
These are distributions of the inferred shared variable at fixed $S$.
Our numerical result is that learning through four-neighbor
communication produces nearly uniform site weights whose local
inference means collectively reproduce magnetization.
}


\section{Conclusion and discussion}
\label{sec:conclusion}

In this article, we have formulated the emergence of macroscopic order through collective predictive coding (CPC). Each site behave as an agent, and it describes its observed spin by a local probability distribution conditioned on a variable common to all the agents. Learning these distributions from configurations, in such a way that the common shared variable can predicts the configurations better, determines which collective patterns the common variable represents. In this manner the shared common variable automatically emerges as functions of the configuration through CPC and learning, and they serve as a macroscopic variable well representing the configurations, thus called collective predictive orders. 

Our CPC construction recovers magnetization and staggered magnetization
in the 2-dimensional classical Ising models, two competing stripe orders, and replica overlap
in the fully connected Sherrington-Kirkpatrick model when each site observes a replica pair, as shown in Sec.~\ref{sec:emergent-order}.
In particular, the emergence of the order parameters is verified at various levels --- the non-necessity of knowing the symmetry structure of the system beforehand, and the automatic detection of the necessary number $K$ of the order parameters. 
For the Ising models, the equilibrium distribution of the collective predictive order yields an emergent Landau potential that evolves from two wells to one on heating. Extrapolating the temperature where its quadratic
coefficient vanishes gives transition temperatures consistent with the known exact value (Sec.~\ref{sec:emergent-potential}).

We also implemented learning through four-neighboring site communication.
Local proposals and acceptance decisions lead from random site weights
to a common variable that closely follows magnetization. The learned site weights recover the uniform ferromagnetic pattern
also found by centralized CPC in Sec.~\ref{sec:s3-ferro},
and the spatial average of the local inference means closely
tracks magnetization.

The novelty of our CPC application is that we regard each atom as an agent. We allow its capability of inference, prediction and communication. The communication among the microscopic agents to better predict the world lets the macroscopic order emerge collectively.  
This concludes that the philosophy of ``Atom-as-an-Agent" principle actually works quite well in classical spin systems for the emergence of order parameters, Landau potentials, and symmetry and its spontaneous breaking. 

Before ending this article, here we argue our novelty concerning the emergence of order parameters by comparing our method to other methods. The emergence of order parameters from microscopic data has been
approached from several complementary directions using machine learning (ML) and neural networks.
We briefly place the present work in this context, for each of those directions, to discuss the difference and the merits of our AaaA principle.

\begin{itemize}

\item \textit{Mean-field theory.}
The conventional Weiss mean-field approximation replaces the influence of neighboring spins by a self-consistent average field determined by the microscopic interactions. Our AaaA learns a collective coordinate directly from spin configurations without requiring these interactions (and thus their Hamiltonian) as input, and uses its measured equilibrium distribution to construct an effective potential that reflects the observed collective fluctuations.

\item \textit{ML methods for phase classification and transition detection.}
Neural networks have been used to classify phases from spin configurations~\cite{Carrasquilla:2017} and electronic wave functions~\cite{Ohtsuki:2016}. Transitions have also been located through classification performance under varied trial phase labels~\cite{vanNieuwenburg:2017} and through the learned weights of a temperature classifier~\cite{Tanaka:2017}. Our AaaA learns a shared variable for each configuration without phase or temperature labels, and tests its physical relevance by predicting unobserved spins. This variable also provides a common description that individual sites can develop through local inference and communication. 
{
AaaA thus provides a physically interpretable application of
unsupervised latent-variable modeling to collective behavior, through distributed interactions forming decentralized Bayesian inference.
}

\item \textit{Learning collective coordinates.}
Principal component analysis (PCA) and variational autoencoders can recover order parameters from spin
configurations~\cite{Wang:2016PCA,Wetzel:2017VAE}. Variational autoencoders have also provided coordinates describing chemical ordering in alloys~\cite{Yin:2021} and identified the phase diagram of the anisotropic Ising model~\cite{Baul:2023}.
{
Probabilistic PCA already formulates the extraction of collective
coordinates as likelihood-based learning of a shared latent-variable
model~\cite{Tipping:1999PPCA}.
}
These methods already provide a mapping from each configuration to a collective coordinate.

{
AaaA uses this established statistical framework to investigate how
individual sites can participate in forming and using a collective
description through decentralized inference based on local communication and individual learning. 
}
Each site's local probabilistic prediction model both contributes to inferring the shared variable and predicts its own spin from that variable.
This gives the collective coordinate a concrete role in interactions among microscopic agents, and allows its formation to be studied through local proposals, acceptance decisions, and learning.

\item \textit{ML methods for interpretable order parameters.}
Support vector machines (SVMs) with tensorial kernels have extracted explicit order-parameter tensors and local constraints in frustrated magnets~\cite{Greitemann:2019Constraints}, including multiple coexisting and competing orders~\cite{Liu:2019}. An SVM classifier for many-body localization has also been related to a physical measure of wavefunction localization~\cite{Zhang:2019MBL}.
The distinction from AaaA lies in the principle used to identify the order parameter. These SVM approaches extract physical observables from learned rules for distinguishing phases.
AaaA learns a shared variable that explains the observations at individual sites, without using phase classification as its
learning objective.
The resulting variable is identified with physical order through its correspondence with known observables and its ability to predict unobserved spins.
Thus, even when the same order parameter is recovered, it emerges as a discriminator between phases in the SVM approaches and as a common explanation of local observations in AaaA.

\item \textit{Coarse graining and effective potentials.}
Coarse-grained variables have been learned by maximizing the information they retain about distant degrees of freedom~\cite{KochJanusz:2018}. This shares with AaaA the aim of identifying collective variables through their statistical relation to other parts of the system;
AaaA additionally gives these variables a role in communication between local agents.
Landau free-energy potentials have also been learned from polarization and strain data~\cite{Pulzone:2025}.
In the present construction, the collective coordinate itself is learned from microscopic configurations, and its measured equilibrium distribution then supplies the effective potential.

\end{itemize}

These approaches offer complementary ways to extract macroscopic physics from microscopic data.
The distinctive emphasis of AaaA is to make each microscopic site a participant in forming the collective description:
each site learns a local probabilistic prediction model, while a shared variable connects its observations to those of other sites.
Our examples show how this construction identifies ordering patterns, supports prediction of unobserved spins, and provides a coordinate for reconstructing an equilibrium free-energy potential.
The communication experiment further demonstrates that an ordering pattern can develop through local proposals, acceptance decisions, and learning.
This gives concrete form to the Atom-as-an-Agent principle: macroscopic order becomes a shared description that microscopic agents learn to use.

Let us discuss possible more applications of our AaaA principle. First, the examples which we studied in this article is just classical systems, and its application to quantum systems is important in dealing with emergence of macroscopic order, since as described in the introduction the emergence of classical physics out of quantum mechanics is one of the major examples of emergent phenomena in physics. It should be noted that in quantum systems more variety of order parameters can be introduced, such as topological phases.
For example, unsupervised learning has classified band topology through continuous deformations connecting Hamiltonians~\cite{Scheurer:2020} and identified topological phase boundaries from experimental ultracold-atom data~\cite{Kaming:2021}. These results extend learning beyond conventional symmetry-breaking order and suggest a direction for AaaA beyond the examples studied here. 
Such an extension would require identifying observations and local probabilistic prediction models through which the shared variable can represent topological information.

A further direction concerns spacetime emergence in holography, as mentioned in the Introduction.
Within a holographic neural-network architecture, bulk metric functions have been learned from boundary response data, including lattice QCD data~\cite{Hashimoto:2018AdSCFT,Hashimoto:2018HolographicQCD}.
These studies motivate asking whether inference and communication among microscopic CPC agents can produce a shared description with a geometric interpretation.
This would extend the present approach from the emergence of order parameters toward the emergence of spacetime.

One of the generalizations of our AaaA protocols is on the architecture of the local communication.
In Sec.~\ref{sec:agentic-communication}, the four-neighbor experiment uses the lattice adjacency. One can modify this communication rule --- how many agents participate in the single communication, and how this communication group is chosen in the learning process. This extension could affect how successfully collective predictive order emerges. For example, if all sites participate in the acceptance decision, the acceptance rate can become very low, making updates of the common variable inefficient. Some other communication architecture can also be suggested from experimental configuration data which often can access only a partial set of sites. It would be interesting to extend our AaaA architecture to the one which reflects the experimental situation.

Finally, let us conclude with a philosophical point of view. Our Atom-as-an-Agent (AaaA) principle may sound odd, as we regard atoms as agents to communicate. However, remember that in these decades the physics principles have been affected by and introduced the viewpoint of information theory. Communication is the most fundamental part of information theory as C.~Shannon himself described \cite{Shannon:1948}. Thus it is rather natural that the functions such as agentic communication and agentic inference are assigned to each atom, with which we reconstruct our physical theory on that novel basis.
Our AaaA principle is a realization of such philosophy.
Interestingly, G.~Parisi, in his account of complex systems,  describes atoms and spins as interacting elementary agents whose collective behavior emerges from simple individual rules~\cite{Parisi:2023Starlings}. 
The AaaA construction pursued here is along Parisi's expression.
Each atom contributes its observation, assesses a proposed common variable, and learns how that variable explains its local state. 
Through the scheme of collective predictive coding, the collective predictive order then acquires meaning as the shared predictive description of these atomic agents.
The realization of this AaaA principle, presented in this article, is one of the goals of Machine Learning Physics. 


\section*{Acknowledgment}
The authors thank the members of the Institute of Generative Science, Kyoto University (Preparatory
Office), for many valuable discussions. 
K.~H.~is grateful to all the members of ``Foundation of Machine Learning Physics" project in Japan for developing and exploring together new perspectives in physics through AI. K.~H.~also like to thank Masayuki Ohzeki for his personal communications.

K.~H.~was supported in part by JSPS KAKENHI Grant No. JP22H05111 and JP22H05115, and by JST,
ARiSE, Japan Grant No. JPMJAR2633. 
T.~T.~was supported in part by JSPS KAKENHI Grant Numbers JP23H04835.
We also acknowledge the use of artificial intelligence tools for their assistance in coding and manuscript preparation.

\let\doi\relax

\end{document}